\documentclass[sigconf,9pt]{acmart}
\usepackage[english]{babel}
\usepackage{color,colortbl,multirow,hyphenat}
\usepackage{tikz,hyperref,xspace,booktabs}
\usepackage{tabularx,enumitem}
\usepackage[labelformat=simple]{subcaption}
\usepackage{multirow}
\usepackage[normalem]{ulem}

\usepackage{booktabs,amsfonts,amssymb,algorithm}
\microtypecontext{spacing=nonfrench}
\newcommand{\parahead}[1]{\vspace*{1ex minus 0.25ex}\noindent %
  {\bfseries #1.}}
\renewcommand{\paragraph}[1]{\vspace{2pt plus 0pt minus 2pt}\noindent{\bfseries #1}}
\newcommand{\parabreak}{\vspace*{1.5ex minus 0.25ex}\noindent}

\newcommand{\ie}{{\itshape i.e.}\xspace}
\newcommand{\eg}{\emph{e.g.}\xspace}

\definecolor{Gray}{gray}{0.9}
\definecolor{LightGreen}{rgb}{0.88,1,0.88}
\definecolor{LightYellow}{rgb}{1,1,0.8}
\definecolor{LightOrange}{rgb}{1,0.85,0.8}
\definecolor{LightRed}{rgb}{1,0.80,0.80}

\newcommand{\systemname}{PBE-CC}
\newcommand{\systemnames}{PBE-CC's}
\newcommand{\shortname}{\systemname}
\newcommand{\shortnames}{\systemnames}

\hypersetup{colorlinks=true, urlcolor=blue,
    citecolor=black, linkcolor=blue, menucolor=black}

\begin{document}
\title[Congestion Control via Endpoint-Centric, Physical-Layer Bandwidth Measurements]{PBE-CC:~Congestion Control via Endpoint-Centric,
\texorpdfstring{\\}{} Physical-Layer Bandwidth Measurements} 
\author{Yaxiong Xie, Fan Yi, Kyle Jamieson}
\affiliation{Department of Computer Science, Princeton University}
\email{{yaxiongx, fanyi, kylej}@cs.princeton.edu}

\renewcommand{\shortauthors}{Yaxiong Xie, Fan Yi, Kyle Jamieson}

\copyrightyear{2020}
\acmYear{2020}
\setcopyright{acmcopyright}\acmConference[SIGCOMM '20]{Annual conference of the ACM Special Interest Group on Data Communication on the applications, technologies, architectures, and protocols for computer communication}{August 10--14, 2020}{Virtual Event, NY, USA}
\acmBooktitle{Annual conference of the ACM Special Interest Group on Data Communication on the applications, technologies, architectures, and protocols for computer communication (SIGCOMM '20), August 10--14, 2020, Virtual Event, NY, USA}
\acmPrice{15.00}
\acmDOI{10.1145/3387514.3405880}
\acmISBN{978-1-4503-7955-7/20/08}

\begin{abstract}
Cellular networks are becoming ever more sophisticated and
overcrowded, imposing the most delay, jitter, and throughput damage to
end-to-end network flows in today's internet.  We therefore argue for
fine-grained mobile endpoint-based wireless measurements to inform a
precise congestion control algorithm through a well-defined API to the
mobile's cellular physical layer.  Our proposed congestion control
algorithm is based on Physical-Layer Bandwidth measurements taken at
the Endpoint (PBE-CC), and captures the latest 5G New Radio
innovations that increase wireless capacity, yet create abrupt rises
and falls in available wireless capacity that the PBE-CC sender can
react to precisely and rapidly. We implement a proof-of-concept
prototype of the PBE measurement module on software-defined radios
and the PBE sender and receiver in C.  An extensive performance
evaluation compares PBE-CC head to head against the  
cellular-aware and wireless-oblivious congestion control protocols
proposed in the research community and in deployment, in mobile and
static mobile scenarios, and over busy and idle networks.  Results
show 6.3\% higher average 
throughput than BBR, while simultaneously reducing
$95^{\mathrm{th}}$ percentile delay by $1.8\times$.

\end{abstract}


%
\maketitle
\thispagestyle{empty}
\section{Introduction}

Most of today's downlink end\hyp{}to\hyp{}end data flows
terminate at a cellular last hop to a mobile endpoint, where
they encounter the most delay, variations in delay, 
loss of their constituent packets,
and limits on their bandwidth.  With the increasingly 
sophisticated design of 
today's and tomorrow's cellular networks in mind, 
this paper argues that it is actually the endpoints 
that are the entities best positioned
to measure the congestion state of an end\hyp{}to\hyp{}end
connection.  We further argue that
the physical layer of the mobile endpoint ought to measure
the congestion state of the 
wireless last hop, and feed these very fine\hyp{}grained 
measurements up to the transport layer
and applications through a well\hyp{}defined API.  
This position follows from three challenges that all congestion
control algorithms face when they operate in today's 
wireless networks.

First, wireless is fundamentally a shared medium.
This means that when a user's flow commences or finishes, other
users associated with the same cell tower experience an
abrupt drop or rise in available wireless capacity
that takes time to be reflected in the flow of 
acknowledgements that today's ack\hyp{}based congestion control 
protocols send back to the sender \cite{Sprout, Verus, BBR}.
Second, in recent years, to achieve high throughput 
and low end\hyp{}to\hyp{}end queuing delay,
senders must now swiftly react to other 
abrupt capacity changes in 
the wireless cellular link that neither the sender nor even 
the cell tower may directly observe.  One reason behind
this change
is that the newest cellular standards, such as 
LTE-Advance~\cite{LTE-Rel-10} and 5G New Radio~\cite{5G} 
aggressively exploit a wireless diversity technique called
\textit{carrier aggregation} to increase wireless capacity,
in which the cellular network aggregates the capacity from two or 
more cellular base stations, making that aggregate
capacity available to a single 
user.  When the cellular network adds or removes base stations 
participating in a user's aggregated capacity, the wireless capacity 
available to each user abruptly changes, accordingly.
Wireless\hyp{}aware congestion control systems centered on a 
single base station, such as Accel\hyp{}Brake Control (ABC)
\cite{ABC, ABC_NSDI} require non\hyp{}trivial 
extensions to share state across cell sites when carrier 
aggregation is enabled. 
Finally, wireless channel quality is inherently 
highly dynamic, due to, \eg, user 
mobility, multipath propagation, and interference from 
neighboring cell towers.  These factors change the 
wireless data rate that a particular user's 
cellular link supports over a time scale known as the
wireless channel \emph{coherence time}, which can be 
as small as milliseconds in the case of 
vehicular\hyp{}speed mobility.
In the event of a handover between cell towers, 
ABC would need to migrate state, 
which is not considered in its design.

Further, the foregoing factors interact, exacerbating 
their effect.  Due to carrier aggregation, an 
end\hyp{}to\hyp{}end connection experiences fluctuation due to
the dynamics of all its aggregated cells, typically 
fewer (two to four) than can offer a smoothing of capacity 
due to statistical multiplexing.

While both base station and
the mobile endpoint are able to observe these
fluctuations, it is only the latter that has fully
up\hyp{}to\hyp{}date state on the wireless connection 
to each and every base station the mobile connects with.
In the current design of the cellular physical layer, however,
mobile users decode only their own channel allocation 
messages, and so cannot track other users' 
channel usage and thus identify idle wireless capacity.

\parabreak{}This paper introduces a new congestion 
control algorithm based on
\emph{\textbf{P}hysical-Layer \textbf{B}andwidth} measurements, 
taken at the mobile \textit{\textbf{E}ndpoint} (\emph{\shortname{}}).
At a high level, \shortname{} is a cross\hyp{}layer
design consisting of two modules.  Our first module comprises
an end\hyp{}to\hyp{}end 
congestion control algorithm loosely based on TCP 
BBR \cite{BBR}, but with senders modified to leverage precise 
congestion control techniques \cite{xcp-sigcomm02} 
when possible.  We harness our end\hyp{}to\hyp{}end congestion control 
to our second module, a wireless physical\hyp{}layer 
capacity measurement module for mobile devices.
Our key innovation is to enable highly accurate capacity 
measurements of the wireless cellular link, which track its 
variations at millisecond\hyp{}timescale granularity, thus 
enabling significantly more precise control over senders'
rates as they attempt to match their sending rate to the 
available wireless capacity, should the bottleneck capacity
be the wireless link itself.  
In the event of an increase in wireless capacity, 
this allows \systemname{} to be rapidly responsive,
detecting the amount of newly\hyp{}emerged idle wireless
capacity and prompting the sender to increase its offered
rate accordingly.  In the event of a decrease in wireless
capacity, this allows \systemname{} senders to rapidly 
quench their sending rate, thus avoiding queuing delays,
as our evaluation demonstrates in drill\hyp{}down
experiments (\S\ref{s:eval}).

Our evaluation shows that most of the time, 
the cellular
link is indeed the bottleneck in the end\hyp{}to\hyp{}end
path, as many congestion control protocols
\cite{Sprout, ABC_NSDI, Verus} assume.  \systemname{} makes
the same initial assumption, leveraging the above
wireless\hyp{}aware precise congestion control functionality 
to more accurately control the sender's pacing, while also
taking into account the number of users sharing the wireless
link, so that each \systemname{} sender can offer a load
that results in an overall\hyp{}fair distribution of wireless
capacity between those users.  
Further refinements allow
\systemname{} senders to gently approach this target
at the connection start, so that other senders
have time to react and adjust accordingly.
However, if \systemname{}
detects an increase in the one\hyp{}way delay of its 
packets that its wireless capacity forecasts do not anticipate,
this triggers a BBR\hyp{}like mechanism to probe the bottleneck
rate based on the pace of acknowledgement packets 
received by the \systemname{} sender.

We have implemented the \systemname{} congestion control 
module in 814 lines of user space C++ code.
Mobile telephone wireless front ends should decode the necessary
frequency bands in order to implement \shortnames{} 
physical\hyp{}layer wireless capacity measurement module,
but their (closed\hyp{}source) firmware does not offer this
functionality, and so we emulate the missing firmware 
functionality using the USRP software\hyp{}defined radio
in our 3,317\hyp{}LoC C implementation.

\begin{table}
\caption{Summary throughput speedup and delay reduction
performance comparison \emph{vs.}~BBR, Verus, and Copa (averaged over 15 idle cellular links and 25 busy links).}
\label{t:summary_perf}
\centering
\begin{tabularx}{\linewidth}{X*4{>{}c}}
\toprule
\multicolumn{2}{c}{} & \multicolumn{1}{c}{}& 
\multicolumn{2}{c}{\textbf{\systemname{} delay reduction}}  \\ \cline{4-5} 
\multicolumn{2}{c}{\multirow{-2}{*}{\textbf{Scheme}}} & 
    \multicolumn{1}{c}{\multirow{-2}{*}{\begin{tabular}[c]{@{}c@{}}\textbf{\systemname{}}\\ 
\textbf{tput.~speedup}\end{tabular}}} & 
\multicolumn{1}{c}{95th.~pctl.} & 
\multicolumn{1}{c}{avg.~delay} \\ \midrule
                        & Busy & 1.04$\times$  & 1.54$\times$  & 1.39$\times$  \\ 
\multirow{-2}{*}{BBR}   & Idle & 1.10$\times$  & 2.07$\times$  & 1.84$\times$  \\ \hline
                        & Busy & 1.25$\times$  & 3.97$\times$  & 2.53$\times$ \\  
\multirow{-2}{*}{Verus} & Idle & \cellcolor{LightGreen} 2.01$\times$  & \cellcolor{LightGreen} 3.44$\times$  & \cellcolor{LightGreen} 2.67$\times$ \\ \hline
                        & Busy & \cellcolor{LightGreen} 10.35$\times$  & \cellcolor{LightOrange} 0.80$\times$  & \cellcolor{LightOrange} 0.80$\times$ \\  
\multirow{-2}{*}{Copa}  & Idle & \cellcolor{LightGreen} 12.94$\times$  & \cellcolor{LightOrange} 0.79$\times$  & \cellcolor{LightOrange} 0.82$\times$ \\ 
\bottomrule
\end{tabularx}
\end{table}

Our performance evaluation uses Pantheon
\cite{Pantheon} to test \systemname{} head\hyp{}to\hyp{}head against
BBR and CUBIC \cite{CUBIC}, leading congestion control algorithms, as
well as recent congestion control algorithms for cellular
\cite{Sprout, Verus}, and other recently\hyp{}proposed algorithms such
as Copa~\cite{Copa}, PCC~\cite{PCC} and PCC-Vivace~\cite{PCC-v}.  Our
experiments begin with measurements of delay and throughput, under
stationary user\hyp{}device conditions, both indoors and outdoors, and
both during busy and quiet hours.  Further experiments evaluate the
same under mobile user\hyp{}device conditions, ``controlled''
competition for the wireless network capacity (that we introduce
ourselves in a known manner), and ``uncontrolled'' competition from 
background traffic of other users at various times of the day.
For each competing
scheme, we report (individually) all throughput and delay 
order statistics,
measured across 100-millisecond time windows, as well as
average case results for these experiments. 
Table~\ref{t:summary_perf} summarizes our performance results:
on average, \systemname{} achieves a $6.3\%$ higher average 
throughput than BBR, while simultaneously reducing
$95^{\mathrm{th}}$ percentile delay by a factor of $1.8\times$ and
average delay by a factor of $1.6\times$.  Against 
Verus, an algorithm specially designed for cellular networks, \systemname{} achieves significant gains in
both throughput and delay reduction, and against the
much\hyp{}slower
Copa, \systemname{} achieves an approximate $11\times$ throughput
improvement while paying a relative 20\% latency penalty.
We also evaluate multi\hyp{}user fairness, RTT fairness and
TCP friendliness of \systemname{} in \S\ref{s:eval:fair}.
 
\begin{figure*}[t]
     \begin{minipage}[htb]{0.22\linewidth}
        \centering
        \includegraphics[width=\textwidth]{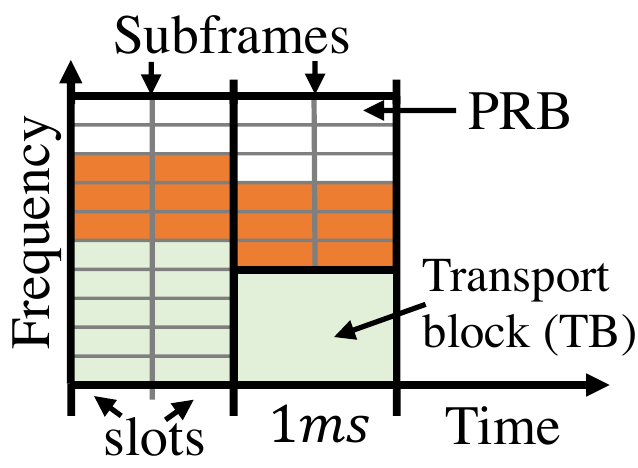}
        \caption{PRBs inside a subframe can be allocated to multiple users. Allocation in two slots are the same (represented using colors).  }
        \label{fig:ofdma}
    \end{minipage}
    \hfill
    \begin{minipage}[htb]{0.37\linewidth}
        \centering
        \includegraphics[width=\linewidth]{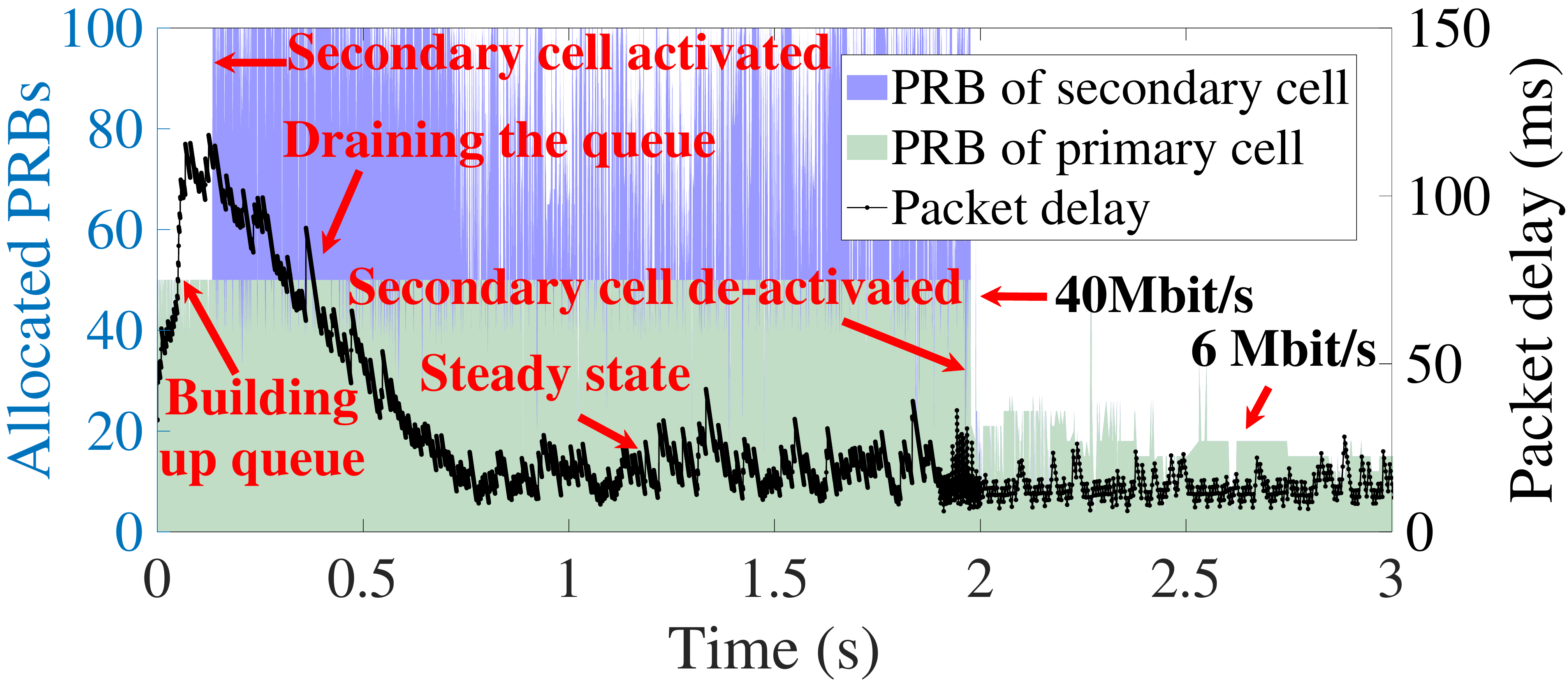}
        \caption{ When the offered load of the server exceeds the maximum capacity of the primary cell, cellular network activates a secondary cell for the mobile user to support the high data rate, and deactivates it if the rate drops.}
        \label{fig:CA_trigger}
    \end{minipage}
    \hfill
    \begin{minipage}[htb]{0.37\linewidth}
        \centering
        \includegraphics[width=0.99\linewidth]{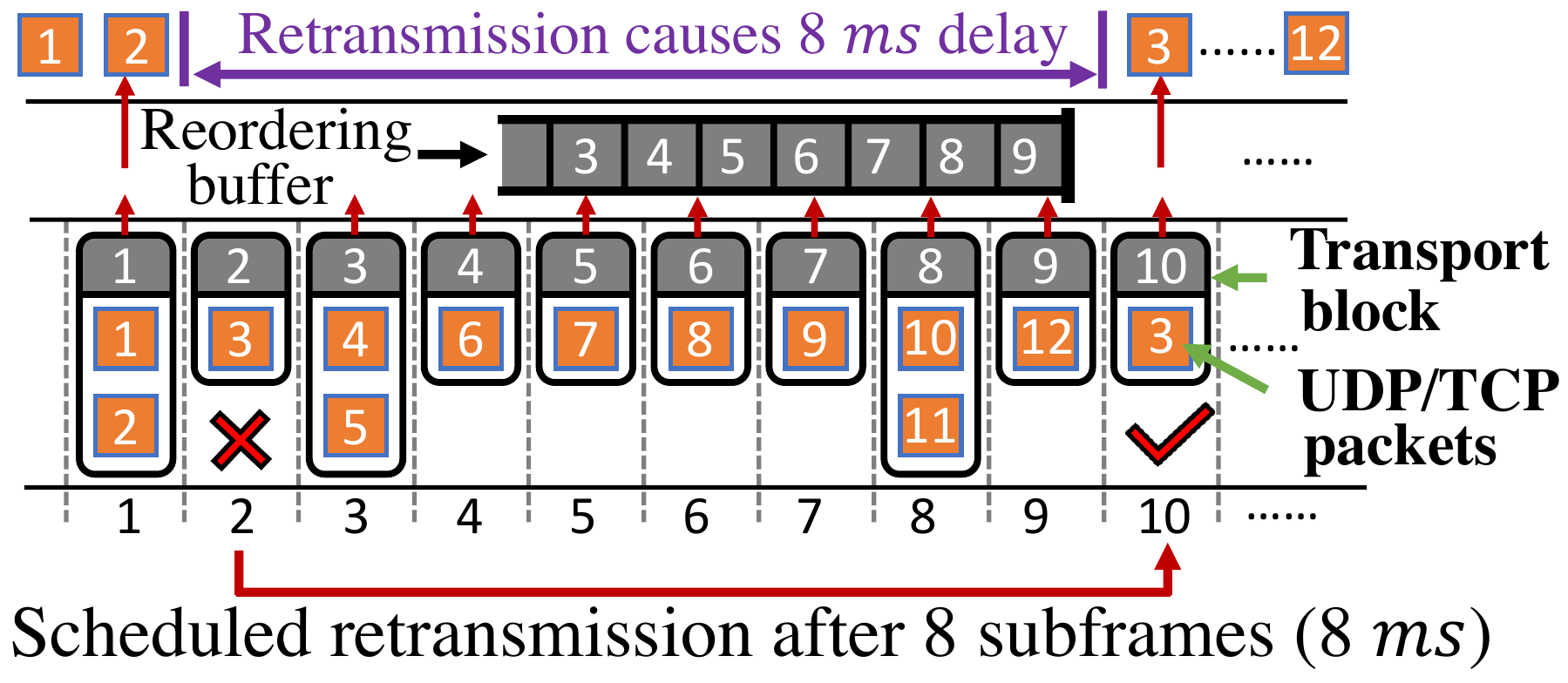}
        \caption{The mobile user buffers all out-of-sequence transport blocks in a reordering buffer until the erroneous block is retransmitted and corrected received (multiple retransmissions is possible), introducing a $8$~ms delay.}
        \label{fig:reorder_buffer}
    \end{minipage}
    \hfill
\end{figure*}

\section{Related Work}
\label{s:related}

\parahead{End\hyp{}to\hyp{}end congestion control}
Loss\hyp{}based algorithms \cite{CUBIC,loss_Reno,delay_loss_Compound,newReno} 
achieve high throughput, but often introduce excessive delay, while
delay\hyp{}based algorithms \cite{delay_Vegas,delay_FAST,Copa} are prone 
to ACK delay, ACK compression, or network jitter, 
and thus often result in network capacity under\hyp{}utilization.
Moreover, it is widely known that these methods achieve poor capacity utilization 
when competing with concurrent loss-based algorithms~\cite{Copa,delay_loss_Compound}.
Other proposals use learned algorithms to optimize specific objective
functions, to generate better congestion 
control actions than human crafted rules~\cite{Remy1,Remy2,PCC,PCC-v,PCP}. 
As we show in our evaluation (\S\ref{s:eval}), 
online learning frequently converges to solutions that result in significant network under\hyp{}utilization. 
BBR~\cite{BBR} targets convergence to Kleinrock's optimal operating
point, \ie, simultaneously maximizing throughput and
minimizing delay, 
based on estimates of bottleneck bandwidth and round trip 
propagation time.
BBR achieves the best performance among all the algorithms we test, 
but still under\hyp{}utilizes the network and introduces excessive 
delay because of its capacity estimates are coarse\hyp{}grained.

\parahead{End\hyp{}to\hyp{}end congestion control for cellular networks}
Some prior work treats the cellular link as a black box and makes use of
throughput, packet delay and loss statistics to infer link capacity~\cite{MaoInDepth}.
Raven~\cite{raven-mobicom18} reduces interactive video latency by sending redundant data 
over multiple paths (Wi\hyp{}Fi and cellular), 
using Multipath TCP \cite{MPTCP}.
PROTEUS~\cite{PROTEUS} collects current throughput, 
loss, and one\hyp{}way delay, using regression trees to forecast 
future network performance. 
PropRate~\cite{RropRate} replaces BBR's periodic bandwidth probing with continuous probing 
that oscillates the send rate around the estimated receive rate using packet size, 
and packet send\fshyp{}receive times. 
Sprout~\cite{Sprout} leverages packet arrival times to infer the uncertain dynamics of 
the network path, forecasting link capacity based on these measurements.
Similarly, ExLL~\cite{ExLL} models the relationship between packet arrival patterns
and cellular bandwidth usage to adjust send rate. 
Instead of attempting to infer the cellular network dynamics, 
Verus~\cite{Verus} tries to learn a delay profile that captures the relationship between 
target send window size and perceived end\hyp{}to\hyp{}end delay. 
Purely relying on end\hyp{}to\hyp{}end statistics, 
above algorithms inevitably suffers from capacity estimation inaccuracies 
and are sensitive to network dynamics, as we have demonstrated (\S\ref{s:delay_thput}). 
\systemname{} delivers superior performance because of its more fine\hyp{}grained capacity estimation,
achieved by directly measuring the wireless channel.

\parahead{Cellular-aware congestion control proposals}
ABC~\cite{ABC,ABC_NSDI} and the Draft IETF Mobile 
Throughput Guidance (MTG) standard~\cite{MTG} propose modifications of 
each mobile base station to explicitly communicate the best rate to the sender,
but do not explicate specifics in the design of the capacity monitor that is 
critical for high performance. 
CQIC~\cite{CQIC} embarks on a cross\hyp{}layer design by
extracting 3G link capacity estimates, but still lacks fine granularity.
piStream~\cite{piStream} and CLAW~\cite{CLAW} formulate a model 
that predicts utilized resource blocks from signal strength measurements. 
CLAW uses this model to speed up web browsing workloads, while 
piStream uses the model for video workloads,  but
the authors' own measurements show that signal strength's 
predictive power is quite limited, while
\systemname{} decodes the control channel metadata directly, 
resulting in precise bandwidth utilization data that are not estimates.

\parahead{Cellular PHY\hyp{}layer monitoring tools}
QXDM~\cite{QXDM} and MobileInsight \cite{MobileInsight}
extract control messages for a single mobile user,
but cannot provide net information on the cell tower's capacity occupancy,
as \systemname{} does.
BurstTracker \cite{BurstTracker} 
locates the bottleneck of an end\hyp{}to\hyp{}end connection.
LTEye~\cite{LTEye} and OWL~\cite{OWL}
decode control messages, but do not 
work with carrier aggregation (\S\ref{s:primer})
and later advanced MIMO standards
as \systemname{} does.  All the foregoing tools
stop short of a congestion control algorithm design.
\section{LTE/5G New Radio Primer}
\label{s:primer}

In this section, we introduce the relevant design of LTE's MAC and physical layer, with a focus on frequency division duplexing (FDD), the mode cellular operators use most widely.
LTE adopts OFDMA, dividing the available wireless frequency bandwidth into 180~KHz chunks and time into 0.5 millisecond slots, as shown in Figure~\ref{fig:ofdma}.
The smallest time-frequency block (180~KHz and 0.5~ms) is called a \textit{physical resource block} (PRB), which is the smallest unit that can be allocated to a user.
LTE groups two slots into a one-millisecond \textit{subframe}. The PRB allocation of two slots inside one subframe is the same. 
The data transmitted over one subframe is called one \textit{transport block} (TB). 
The size of one TB varies, depending on the number of allocated PRBs and 
the wireless physical data rate of the user. 
The base station informs the mobile user of its bandwidth allocation 
(the amount and position of allocated PRBs) and wireless bit rate, 
including the modulation and coding scheme (MCS) and the number of 
spatial streams, through a control message transmitted over a 
\textit{physical control channel}~\cite{TS212}. 
A mobile user decodes the control message of a subframe before decoding the TB inside it. 

\begin{figure*}[htp]
\centering
\includegraphics[width=0.7\linewidth]{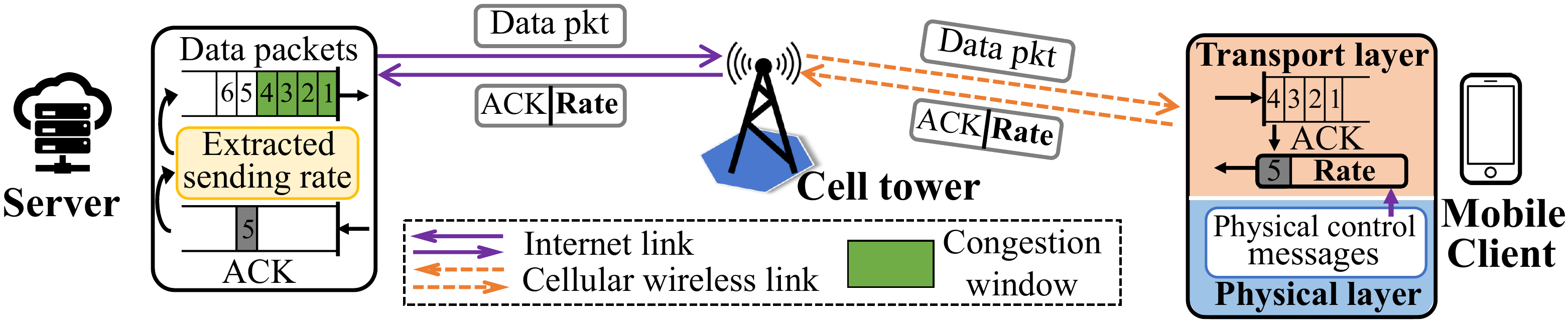}
\caption{An overview of \systemname{} congestion control. 
The mobile clients decode the cellular control channel, which contains detailed information about the base station's available wireless capacity.
\systemname{} senders control their send rate based on the estimated bottleneck capacity that the mobile user explicitly sends back, or based on the presence of ACKs from the receiver.
}
\label{fig:sys_arch}
\end{figure*}

\parahead{Carrier aggregation}
By default, the base station delivers data to a mobile user via a primary \textit{component carrier} (CC), or primary cell.
When there is a huge amount of data to be delivered to the user, the base station activates a secondary cell to add capacity.
The cellular network maintains a list of aggregated cells for each user and will activate them sequentially if necessary.
The aggregated cells are deactivated if and when the user does not utilize the extra capacity.
An example of the carrier activation and deactivation process is shown in Figure~\ref{fig:CA_trigger}. 
A sender first sends data to a mobile user with a fixed offered load of 40~Mbits for two seconds, 
which exceeds the maximum capacity of the primary cell, 
so it causes packet buffering at this cell,\footnote{
We note that packet buffering at the base station is not a prerequisite for activating secondary cells.
The cellular network activates another cell for a user as long as such a user is consuming a large fraction of the bandwidth of the serving cell(s).}
even when all the bandwidth are allocated for this user.
The cellular network detects such a high\hyp{}data\hyp{}rate user and activates a secondary cell to help deliver the data to this user, at 0.13 seconds.
Since 40~Mbit/s is below the aggregated capacity of the primary and secondary cell, 
the cellular network drains the built queue within 0.6 seconds, as shown in Figure~\ref{fig:CA_trigger}. 
The sender reduces its sending rate to 6~Mbit/s, which is below the capacity of the primary cell, 
so the secondary cell is deactivated.

\parahead{Cellular retransmission and reordering}
The cellular network retransmits an erroneous transport block after eight subframes (milliseconds) of the original transmission, as shown in Figure~\ref{fig:reorder_buffer}.
To guarantee in order delivery, the mobile user buffers all the transport blocks received in subframes between the original transmission 
and retransmission of the erroneous transport block (supposing they are received correctly) in a reordering buffer.
When the retransmission succeeds, the mobile user report all the buffered transport blocks together with the retransmitted transport block to upper layers 
where the transport layer packets inside the transport blocks are extracted.
As a result, the retransmission introduces a eight millisecond delay to the transport layer packets inside the erroneous transport block 
and the buffering and reordering operations at the receiver side introduces a decreasing delay (from seven to zero milliseconds) to the packets inside the following transport blocks.
If the retransmission fails, the cellular network repeats
the retransmission at most three times, introducing a latency penalty equal to a 
multiple (smaller than three) of eight milliseconds.

\section{Design}
\label{s:design}

\systemname{} is a rate based, end\hyp{}to\hyp{}end congestion control algorithm 
for flows traversing cellular networks and terminating at mobile devices.
\textbf{\systemname{} mobile clients} decode the cellular physical
control channel, which contains detailed information about the base 
station's available wireless capacity.
From this, the mobile user is able to estimate this quantity 
accurately, at millisecond time granularity.
Depending on the location of the bottleneck link, 
\textbf{\systemname{} senders} control their send rate based 
on the estimated bottleneck capacity that the mobile user explicitly sends back, 
or based on the presence of ACKs from the receiver, as shown in Figure~\ref{fig:sys_arch}.
Using its fine\hyp{}grained capacity estimates, 
when the bottleneck is the wireless hop, \systemname{} can 
immediately increase its send rate to grab new available 
capacity without causing any congestion, 
and decrease its send rate accordingly, if competition with other mobile 
users or the wireless channel reduces wireless capacity.

As traffic patterns are highly dynamic,
end\hyp{}to\hyp{}end connections face two possible network states, 
depending on the relative capacities of the bottleneck link 
in the Internet, and the cellular link.
Most of the time, connections are in what we term a
\textit{wireless\hyp{}bottleneck} state where the wireless 
cellular link is the bottleneck of the whole end\hyp{}to\hyp{}end
connection.
In this state, the \systemname{} mobile user can estimate and 
track the bottleneck capacity of the whole connection at millisecond 
granularity by decoding the cellular physical control channel
(\S\ref{s:bandEst_Cellular}). 
The \systemname{} sender matches its send rate with the bottleneck 
capacity that the mobile user explicitly feeds back, almost exactly 
utilizing capacity and at the same time 
causing minimal packet buffering in the network. 
On the other hand, the connection is in an
\textit{Internet\hyp{}bottleneck} state if the 
capacity of the Internet bottleneck 
is smaller than the capacity of the wireless cellular link.
\systemname{} then switches to a cellular\hyp{}tailored 
BBR\hyp{}like congestion control strategy, to compete fairly
with other flows that share the Internet bottleneck for a fair 
share of the bottleneck capacity (\S\ref{s:bandEst_Internet}).
\systemname{} tracks possible changes in these two states, 
controlling the sender's actions accordingly. 

Kleinrock has proven that 
the operating point---maximizing delivered bandwidth while minimizing delay---is optimal for 
both individual connections and the network as a whole~\cite{kpoint1,kpoint2}. 
The operating point is characterized by the insight that one should keep the pipe only just full.
\systemname{} shares the same goal as BBR, which 
is to fill the pipe and minimize the buffering inside the network.
\systemname{} limits the amount of inflight data to the 
bandwidth\hyp{}delay product (BDP) calculated using estimated
round\hyp{}trip propagation time $\mathrm{RTprop}$  and bottleneck capacity
with a congestion window, as shown in Figure~\ref{fig:sys_arch},
so \systemname{} senders often do not send excessive packets 
even when the feedback from mobile user is delayed,
minimizing queuing in the network, for very low latency,
as our experimental evaluation later demonstrates (\S\ref{s:eval}).

\subsection{Connection Start: Linear Rate Increase}
\label{s:fair_share}

On connection start, a \systemname{} sender executes
a \emph{linear rate increase} in order to 
approach a fair\hyp{}share of the bottleneck capacity.
By decoding the control channel, each \systemname{} user 
knows the number of other users sharing the cell bandwidth, 
as shown in Figure~\ref{fig:PRB_allocation}.
\systemname{} therefore calculates expected fair\hyp{}share 
bandwidth (in units of PRBs) $P_{\mathrm{exp}}$ using the
total PRBs available in the cell $P_{\mathrm{cell}}$ and 
the number of active users $N$ (including the mobile itself):
\begin{equation}
P_{\mathrm{exp}} =  P_{\mathrm{cell}} / N.
\label{eqn:exp_prb}
\end{equation}
The user then estimates its expected 
fair\hyp{}share send rate $C_{f}$ (in units
of bits per subframe) as:
\begin{equation}
C_{f} = R_w \cdot P_{\mathrm{exp}},
\label{eqn:fair_rate}
\end{equation}
where $R_w$ is the wireless physical data rate (with units
of bits per PRB) calculated using the number of spatial 
streams together with the coding and modulation rate
for each stream.

The \systemname{} sender linearly increases its send rate from zero to 
the fair\hyp{}share send rate $C_{f}$ in three RTTs. 
The mobile user updates $C_f$ every millisecond, 
and sends the calculated rate back to the server in each acknowledgement.
\systemnames{} linear increase prevents bursty traffic 
and leaves time for 
the cell tower and the other users sharing that tower 
to react to the increased traffic.
The cell tower reacts to the mobile user's increasing 
send rate by proportionally
allocating more bandwidth, 
which results in less bandwidth allocated 
to other users.
Another \systemname{} user immediately detects such a 
decrease in its allocated bandwidth and signals its 
sender to lower its send rate accordingly. 
Eventually, all \systemnames{} users tend to achieve 
equilibrium with an equally\hyp{}shared bandwidth.
When two or more component carriers are active during the
fair\hyp{}share approaching state, 
we calculate
target send rate separately for each aggregated cell,
and sum them up as $C_f$.
When more carriers are activated during 
congestion avoidance (\S\ref{s:ca}), \systemname{} restarts this
fair\hyp{}share approaching process.

The user ends linear rate increase and enters congestion 
avoidance when it achieves its fair\hyp{}share sending rate $C_{f}$.
If the bottleneck of the connection is inside the Internet, rate
$C_{f}$ is not achievable,
so the achieved throughput at the cell tower 
stays at a rate below $C_{f}$ and end\hyp{}to\hyp{}end
packet delay increases with increasing sender 
offered load.  
When the mobile user detects that the receiving rate stops
increasing for one $\mathrm{RTprop}$, while the oneway packet delay increases 
monotonically with an increasing offered load,
it also ends the linear rate increase phase and switches to 
our cellular\hyp{}tailored BBR to handle congestion 
in the Internet (\S\ref{s:bandEst_Internet}). 

\subsection{Steady State: Congestion Avoidance}
\label{s:ca}

We now present the design of \systemnames{} congestion avoidance
algorithm.  When the connection is in the wireless
bottleneck state, \systemname{} senders match their
send rate to estimated wireless capacity (\S\ref{s:bandEst_Cellular}).
Similar to connection startup, \systemname{} identifies a possible 
transition from a wireless\hyp{}bottleneck to Internet\hyp{}bottleneck
state (\S\ref{S:track_alternation}), 
and if this happens, switches to to a cellular\hyp{}tailored 
BBR (\S\ref{s:bandEst_Internet}) to compete fairly with flows
at the bottleneck.

\subsubsection{Wireless Bottleneck State}
\label{s:bandEst_Cellular}

Here a \systemname{} mobile user estimates the available 
cellular wireless capacity $C_p$ (in units of bits per subframe) as
\begin{equation}
C_p = \sum^{N_{\mathrm{cell}}}_{i=1}\left( R_{w,i}  \cdot \left(P_{a,i} + \frac{1}{N_{i}}P_{\mathrm{idle},i}\right)\right)
\label{eqn:capacity}
\end{equation}
where $N_{\mathrm{cell}}$ is the number of activated cells for this user, 
$P_{a,i}$ is the number of PRBs allocated for this user in the $i$th cell,
$N_{i}$ is the number of mobile users in the $i$th cell,
and $P_{\mathrm{idle},i}$ represents the number of idle PRBs 
in the $i$th cell:
\begin{equation}
P_{\mathrm{idle},i} = P_{\mathrm{cell},i} - \sum_{j=1}^{N_{i}}P^j_{a,i}
\label{eqn:idle_prb}
\end{equation}
where $P^j_{a,i}$ represents the allocated PRB for user $j$ of the $i$th cell. 
To smooth the estimation results, we average the calculated 
$R_{w,i}$, $P_{\mathrm{idle},i}$ and $P_{a,i}$ from the most 
recent $\mathrm{RTprop}$ subframes 
(\eg, we average the above parameters over the most recent 
40 subframes if the connection RTT is $40$~ms).
\begin{figure}[tbh]
\centering
\includegraphics[width=0.9\linewidth]{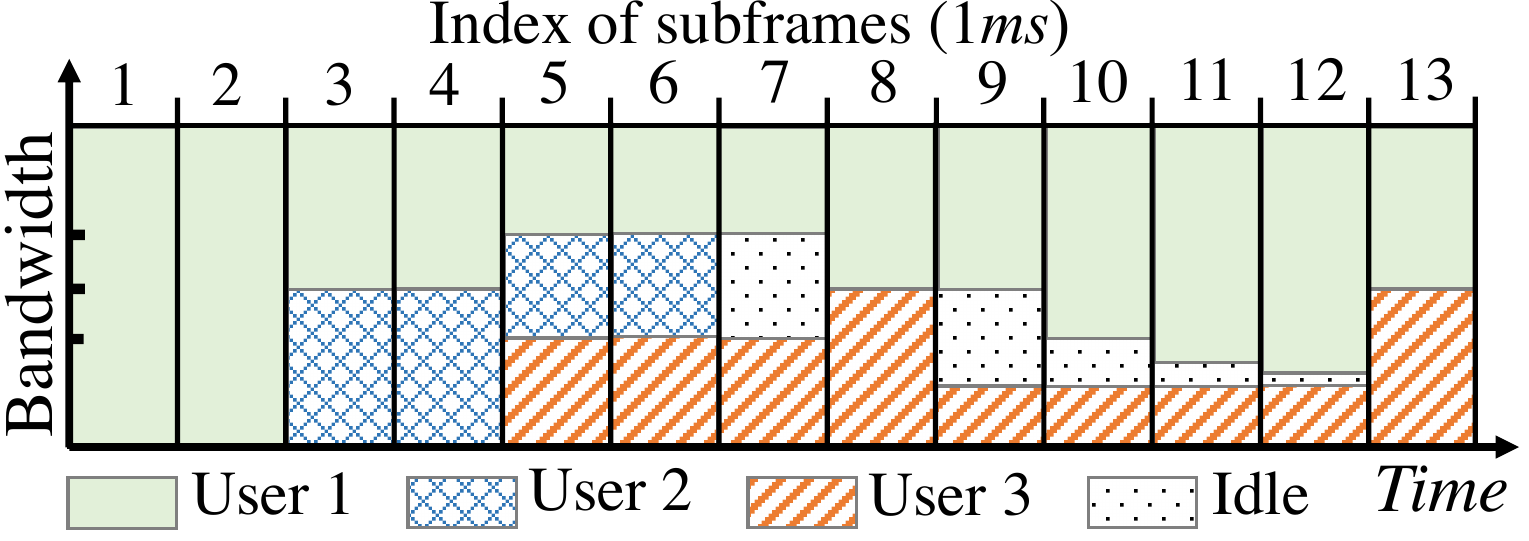}
\caption{One mobile user tracks the number of PRBs allocated for itself, for other mobile users and that are idle.}
\label{fig:PRB_allocation}
\end{figure}

To interpret estimated capacity $C_p$, we consider 
each component of Eqn.~\ref{eqn:capacity}.
First, the wireless physical layer data rate $R_w$ enables the mobile 
user to track capacity variations caused by varying 
channel quality. Second, the mobile user reacts to the appearance of 
new users by tracking the number of PRBs allocated for itself ($P_a$). 
For example, as shown in Figure~\ref{fig:PRB_allocation}, $P_a$ 
for User~1 decreases when a new user, \ie, User~2, starts receiving
traffic.  On detection of fewer allocated PRBs, User~1's sender 
lowers its send rate to match the decreasing capacity 
estimated using Eqn.~\ref{eqn:capacity}.

When idle PRBs $P_{\mathrm{idle}}$ appear in a cell for a connection that is 
wirelessly bottlenecked, 
all \systemname{} clients immediately detect them by checking 
the decoded control message, 
and inform their senders to increase their rates to grab a fair\hyp{}share
portion of the idle PRBs, \ie,  $P_{\mathrm{idle}}/N$.
This may happen in several cases: 
first, idle PRBs appear when a sender finishes a flow.
As shown in the example of Figure~\ref{fig:PRB_allocation}, 
after User~2's flow finishes 
in subframe six, Users~1 and~3 immediately observe idle 
PRBs in subframe seven and then share the available PRBs equally
in subframe eight.
Second, idle PRBs also appear when the data rate of a
user's flow decreases, \eg, Subframe~9 in Figure~\ref{fig:PRB_allocation},
which could be caused by, \eg, congestion in the Internet, 
the application itself,
or a shift of traffic from one cell to another aggregated cell 
by the cellular network.
In this case, all other users immediately detect and occupy their fair share of the newly\hyp{}idle PRBs. 
Other users share $1/N$ of the idle PRBs with User~3, 
whose data rate is limited and thus is not able to grab more PRBs.
As a result, if we define the number of idle PRBs in 
Subframe~9 as $P^\prime$, 
there will be $P^\prime/N$ left idle in Subframe~10. 
Similarly, other users detect these idle PRBs in Subframe~11, but 
still only occupy their fair share portion, 
so $P^\prime/N^2$ will be left idle in Subframe~12.
The network converges to a state where all 
other users other than the User~2 grab all the idle bandwidth.

\begin{figure}[tbh]
    \centering
    \begin{subfigure}[b]{0.49\linewidth}
        \centering
        \includegraphics[width=0.99\textwidth]{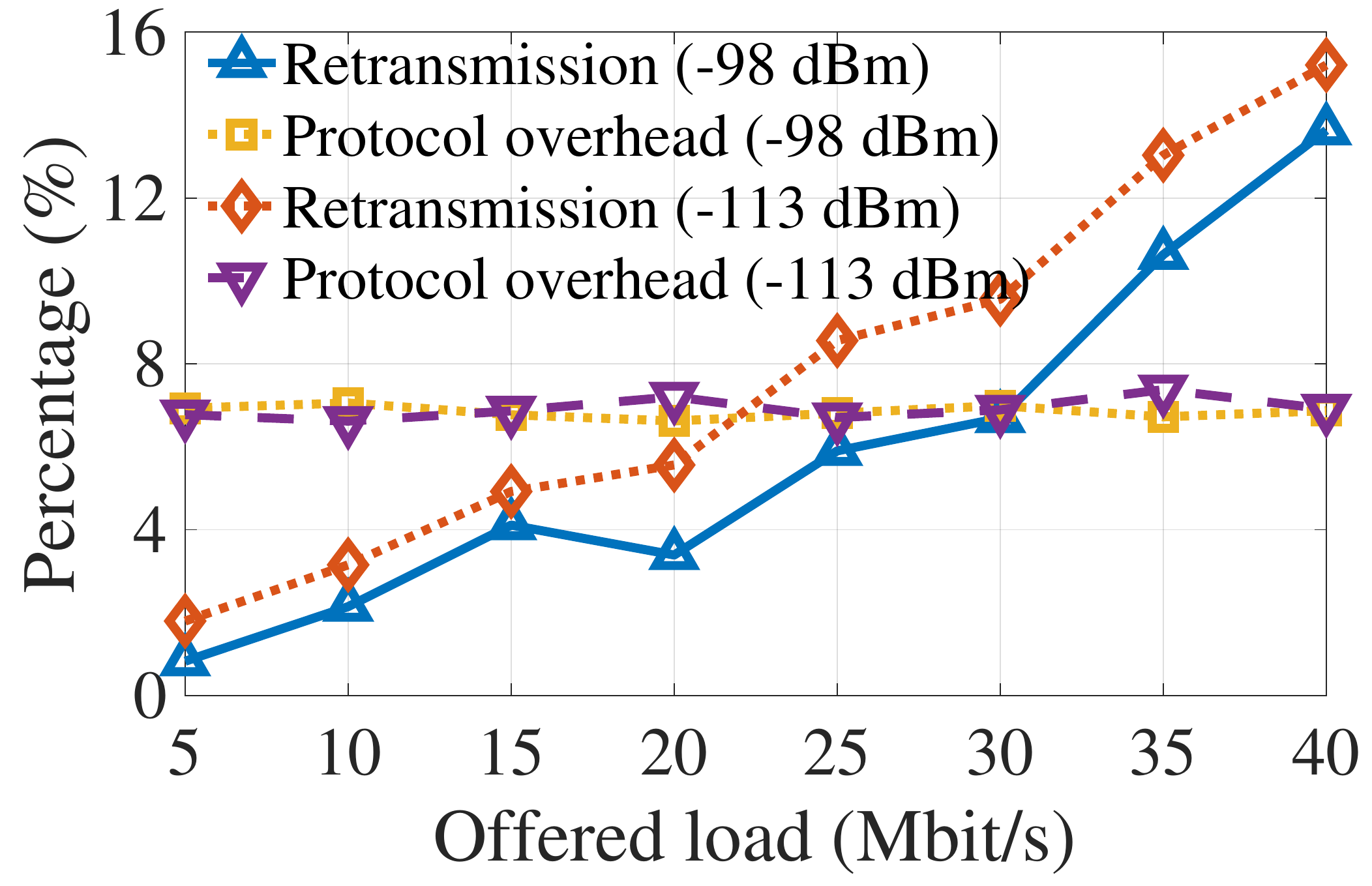}
        \caption{Percentage of overhead.}
        \label{fig:overhead}
    \end{subfigure}
    \hfill
    \begin{subfigure}[b]{0.49\linewidth}
        \centering
        \includegraphics[width=0.99\textwidth]{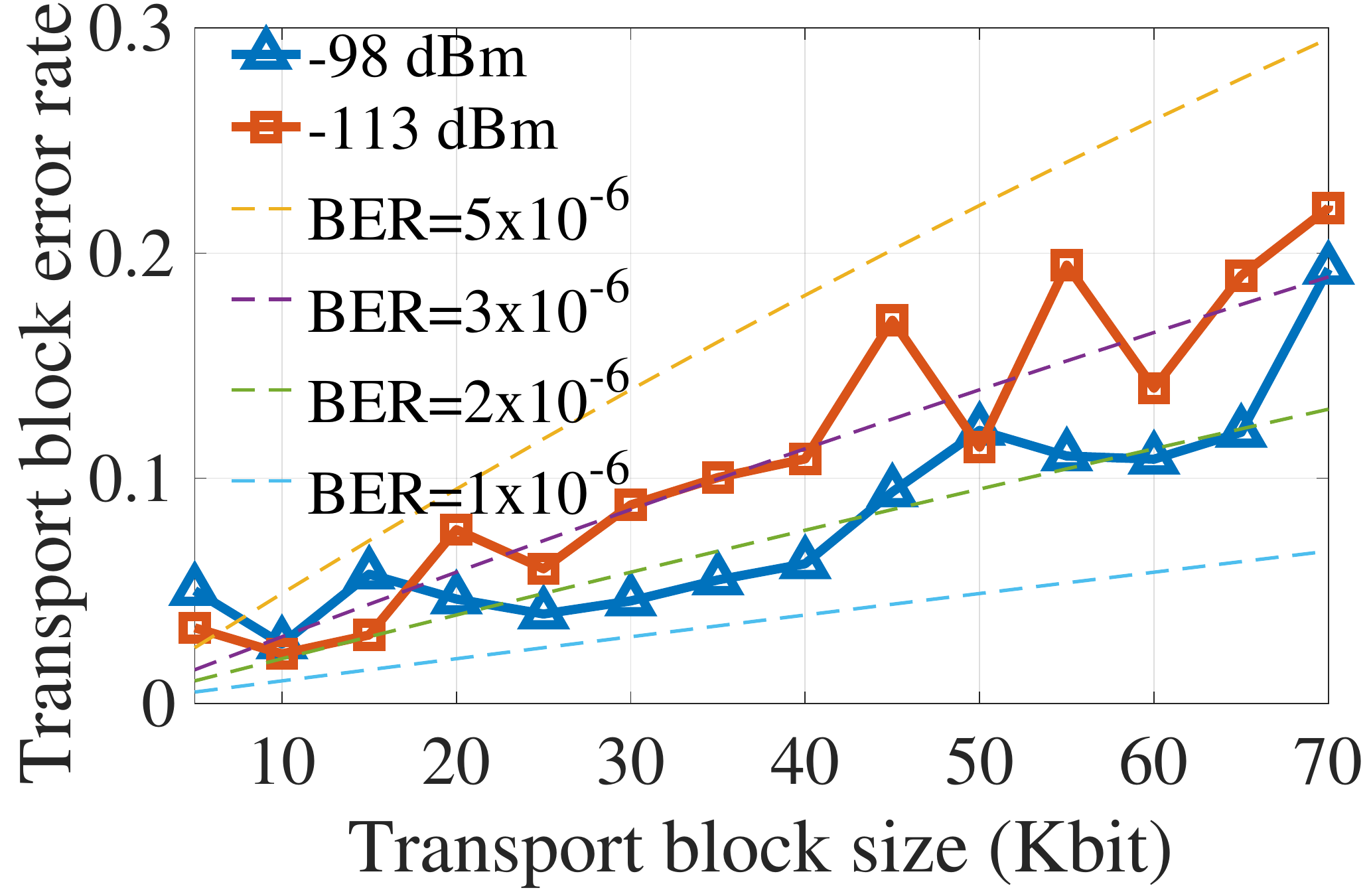}
        \caption{Block error rate.}
        \label{fig:TB_err_TB_size}
    \end{subfigure}
    \caption{The percentage of capacity used for transport block retransmission and transmission of protocol overhead is given in (a).
    The relationship between transport block error rate and transport block size is given in (b).}
    \label{fig:TB_error}
\end{figure}

\paragraph{Cross-layer bit rate translation}
The capacities $C_{f}$ and $C_p$ (Eqns.~\ref{eqn:fair_rate} and~\ref{eqn:capacity})
are wireless physical\hyp{}layer capacities differing from
transport\hyp{}layer data rates due to MAC\hyp{}layer retransmissions 
and (constant) protocol header overhead.  \systemname{} therefore needs to 
transform the estimated physical\hyp{}layer capacity 
$C_p$ to a transport layer goodput $C_t$, and feedback $C_t$ back to the server 
to set its send rate.  The cell indicates a retransmitted transport block 
using a \textit{new-data-indicator}, so we can separately measure retransmission 
overhead and protocol overhead.  Figure~\ref{fig:overhead} plots the measured 
overhead at two different locations and 
varying sender offered loads.
The probability of a TB error determines retransmission overhead:
if the bit error rate (BER) of each bit inside one 
TB is $p$ and bit errors are \textit{i.i.d.}, 
the TB error rate is $1-\left(1-p\right)^L$, where $L$ 
is the TB size.
We plot in Figure~\ref{fig:TB_err_TB_size} theoretical TB error rate (for 
$p=5\times10^{-6}$, $3\times10^{-6}$, and $1\times10^{-6}$)
and empirical TB error rate,
noting a good fit between experimental data and 
theory. Based on these results, \systemname{} models the 
relationship between $C_p$ and $C_t$ as
\begin{equation}
C_p = C_t + C_t 
    \cdot \left(1-\left(1-p\right)^L\right) + \gamma \cdot C_p
\label{eqn:cap}
\end{equation} 
where $\gamma=6.8\%$ is the protocol overhead.  
When one user takes its \systemname{}\hyp{}allocated fair\hyp{}share 
capacity (Eqn.~\ref{eqn:capacity}), the TB size $L$ (number of bits in 
one subframe, \ie, $10^{-3}$~s), is $L = C_t \cdot 10^{-3}$. 
We estimate $p$ using measured \textit{signal to interference noise 
ratio} (SINR), then by solving Eq.~\ref{eqn:cap} given a measured 
physical layer capacity $C_p$, we estimate transport layer goodput $C_t$.
To speed up the calculation, \systemname{} uses a look\hyp{}up table to store the transformation.

\begin{figure}[t]
    \centering
    \begin{subfigure}[b]{0.45\linewidth}
        \centering
        \includegraphics[width=0.99\textwidth]{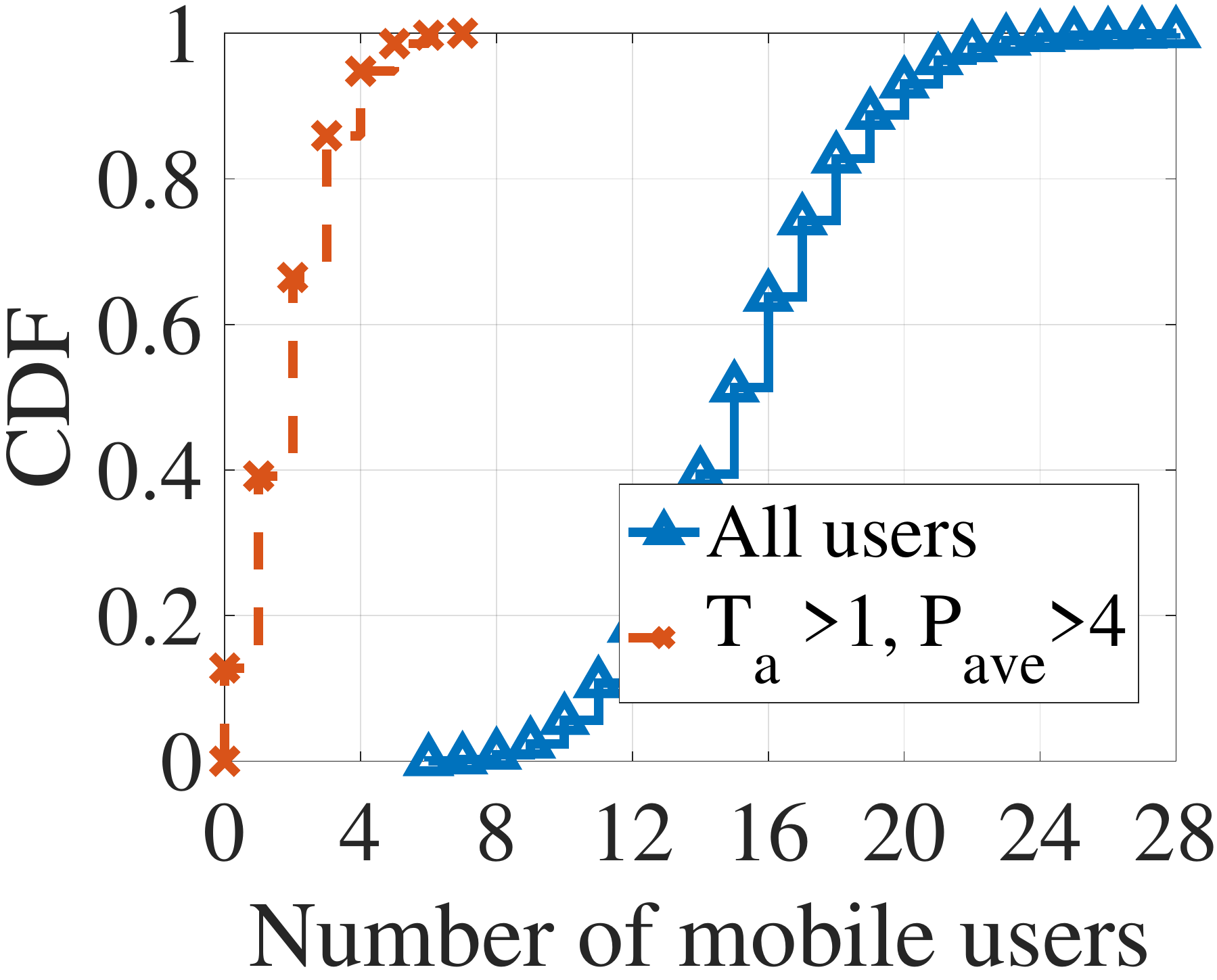}
        \caption{Number of active users.}
        \label{fig:nof_users}
    \end{subfigure}
    \hfill
    \begin{subfigure}[b]{0.52\linewidth}
        \centering
        \includegraphics[width=0.99\textwidth]{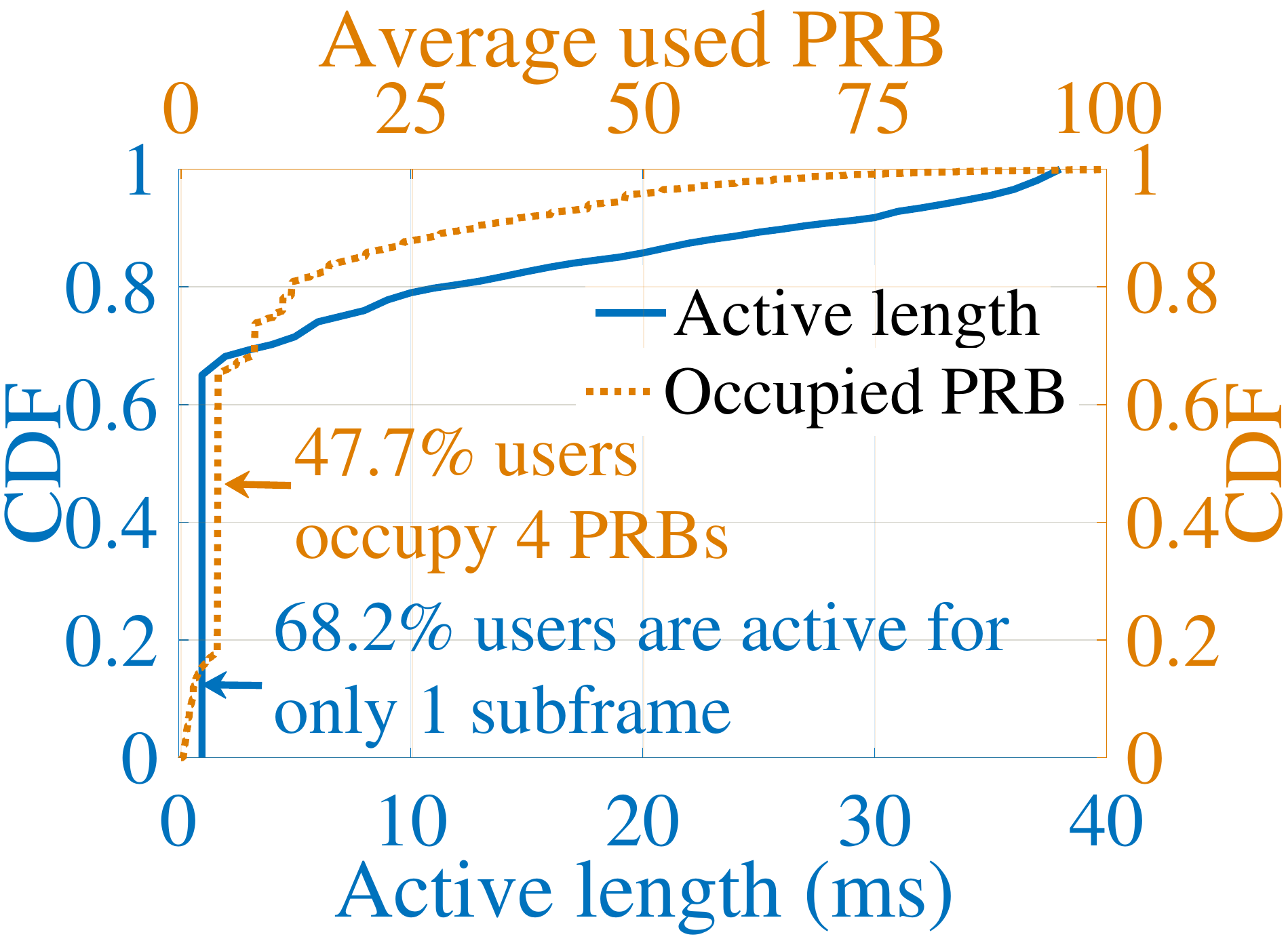}
        \caption{Measured $T_a$ and $P_{\mathrm{ave}}$.}
        \label{fig:active_len_PRB}
    \end{subfigure}
    \caption{Number of mobile users exchanging data with the base 
    station \textbf{(a)}, and 
    activity length $T_a$ and average consumed 
    PRBs $P_{\mathrm{ave}}$ of each detected mobile user \textbf{(b).}}
    \label{fig:active_user}
\end{figure}

\parahead{Handling control traffic}
\systemname{} aims to fairly share wireless bandwidth between all active users,
but our experimental results shows that significant amount of detected users are active not for data, 
but rather to update network 
parameters shared by both base station and 
mobile, \eg, the periods of various 
timers, list of aggregated cells, and many pricing and
security\hyp{}related parameters.
Because of such users, the number of detected active users at each time point could be large. 
For example, we plot the distribution of the number of detected active users in a 40~ms interval, across a 5 hour interval, 
measured from a busy cell tower, in Figure~\ref{fig:active_len_PRB}. 
On average, we observe on average 15.8 and maximum 28 active users, in those 40~ms interval.
\systemname{} excludes those users in its fair\hyp{}share
capacity calculation, reverting to the cell tower to 
allocate small amounts of bandwidth for these users
and then reacting to that allocation by tracking the decrease of 
allocated bandwidth ($P_a$ in Eqn.~\ref{eqn:capacity}) 
and lowering send rate by that amount.
Our key observation is that the control traffic occupies a small number of PRBs and only active for small amount of time.
We plot the distribution of the average occupied PRBs and active time (subframes) of all detected active users in Figure~\ref{fig:active_len_PRB}.
We see that 68.2\% of users occupies exactly four PRBs and is active for exactly one subframe, 
among which 95\% of users are receiving control traffic from the base station.
Therefore, the \systemname{} monitor filters users that are only active for parameter updating, 
based on thresholding the active time duration (subframes) and allocated bandwidth (PRBs) $(T_a>1, P_{\mathrm{a}}>4)$,
after which the number of detected active users
decreases significantly---the average number of detected user inside a 40~ms interval decreases from 15 to 1.3, 
and we only observe at most
seven active users competing for the bandwidth simultaneously, as shown in Figure~\ref{fig:nof_users}.
We set the $N$ in Eqns.~\ref{eqn:fair_rate} 
and~\ref{eqn:capacity} to the number of active users we detect
after applying the threshold. 
The calculation of idle PRBs in Eqn.~\ref{eqn:idle_prb}, however, 
takes every identified user into account. 

\subsubsection{Switching between Bottleneck States}
\label{S:track_alternation}

When sender offered load exceeds the capacity of the Internet bottleneck,
packet queuing induces \systemname{} to switch from the
wireless bottleneck state to the Internet bottleneck state.
\systemname{} triggers a switch 
when the instantaneous one\hyp{}way packet delay exceeds 
a threshold.
Theoretically, we should set the threshold to the one way propagation delay between the server and clients ($D_{\mathrm{th}}=D_{\mathrm{prop}}$).
\systemname{} estimates $D_{\mathrm{prop}}$ as the minimum delay observed in a 10\hyp{}second window,
evoking BBR's round\hyp{}trip propagation delay estimation method.
\systemname{} also updates the true $D_{\mathrm{prop}}$ by draining 
the buffer as BBR does, if estimated packet delay 
maintains constant for 10 seconds.

\begin{figure}[tbh]
    \centering
    \begin{subfigure}[b]{0.35\linewidth}
        \centering
        \includegraphics[width=0.99\textwidth]{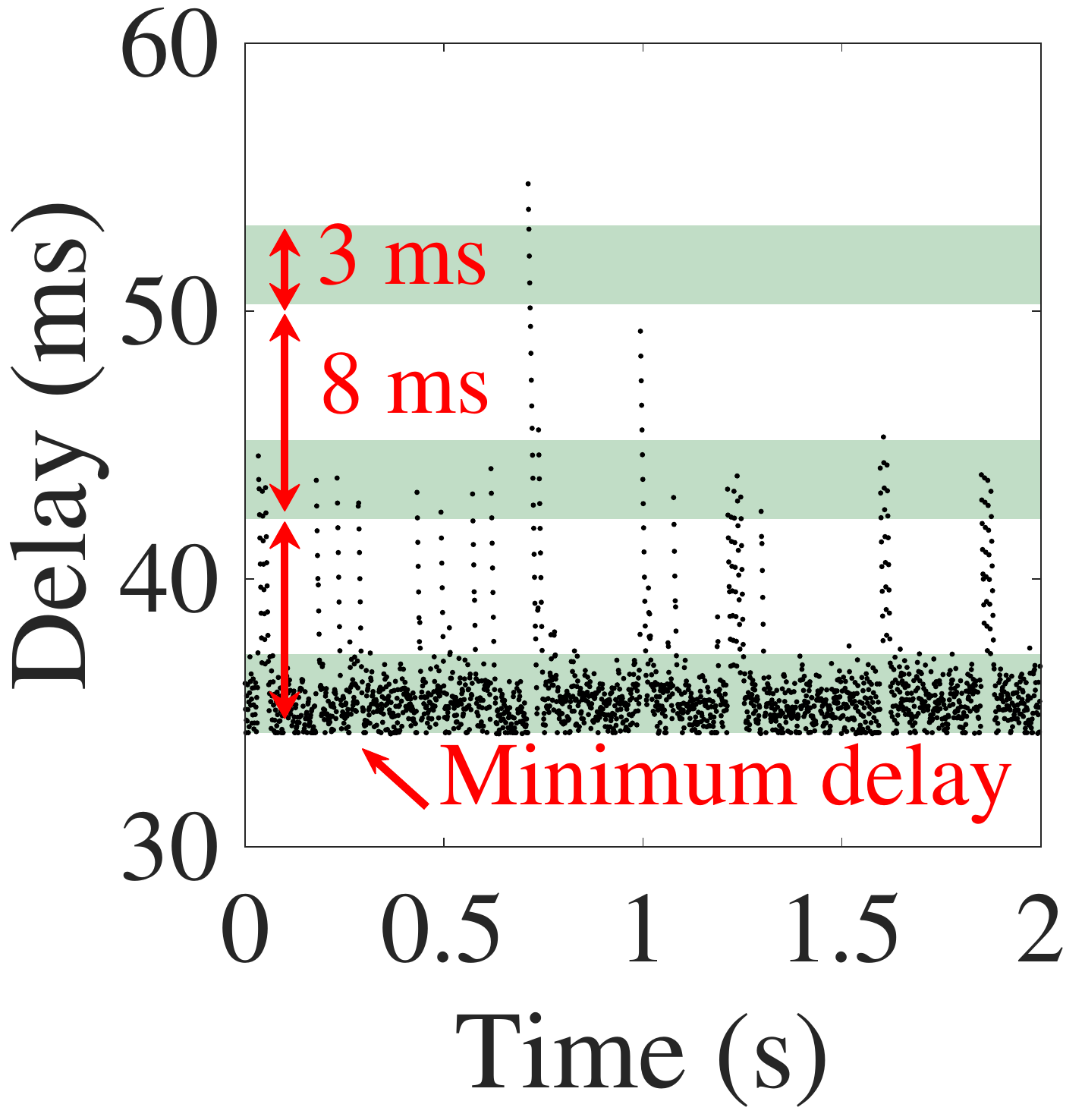}
        \caption{6~Mbit/s.}
        \label{fig:minDelay_a}
    \end{subfigure}
    \hfill
    \begin{subfigure}[b]{0.31\linewidth}
        \centering
        \includegraphics[width=0.99\textwidth]{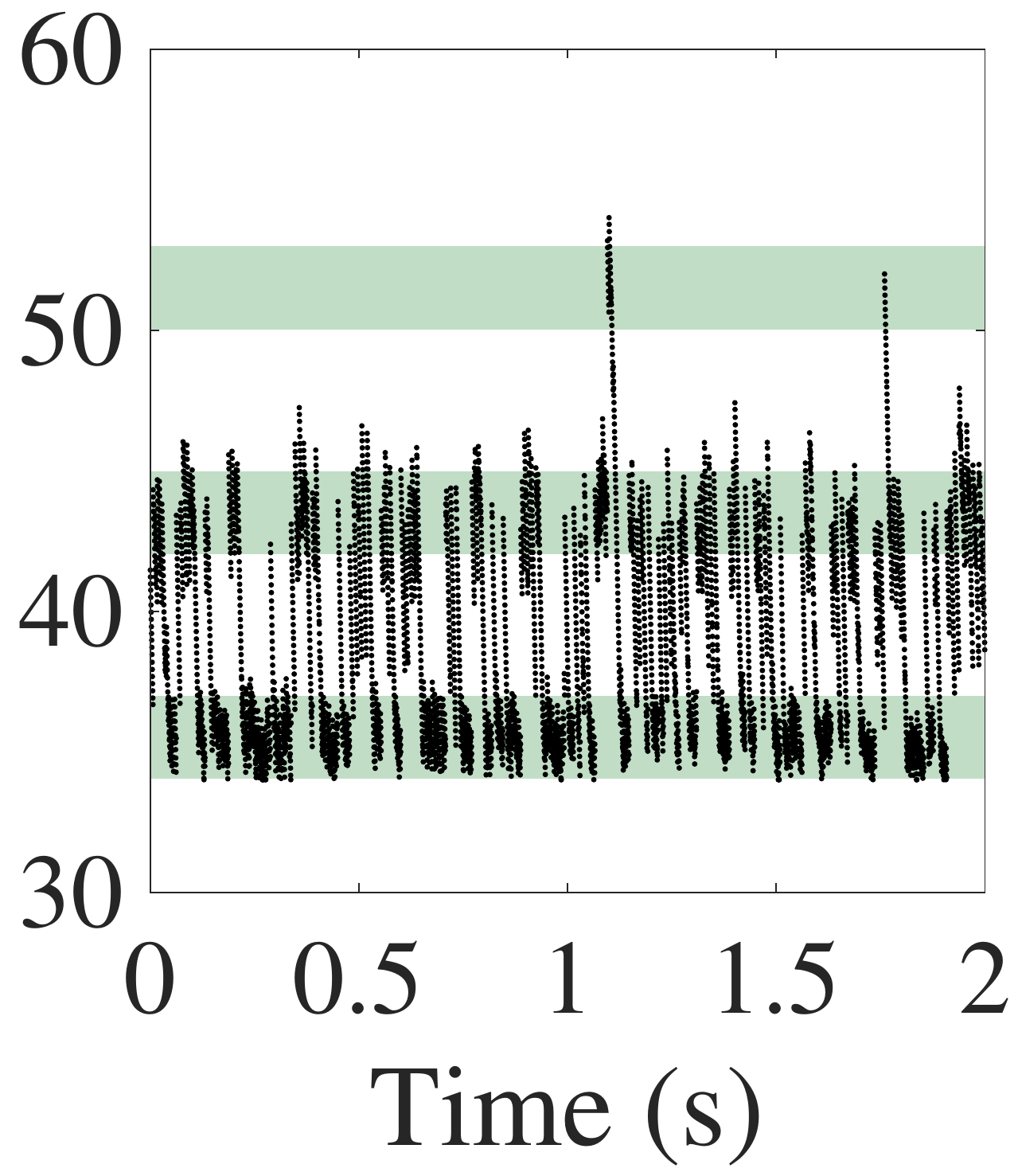}
        \caption{24~Mbit/s.}
        \label{fig:minDelay_b}
    \end{subfigure}
    \hfill
    \begin{subfigure}[b]{0.31\linewidth}
        \centering
        \includegraphics[width=0.99\textwidth]{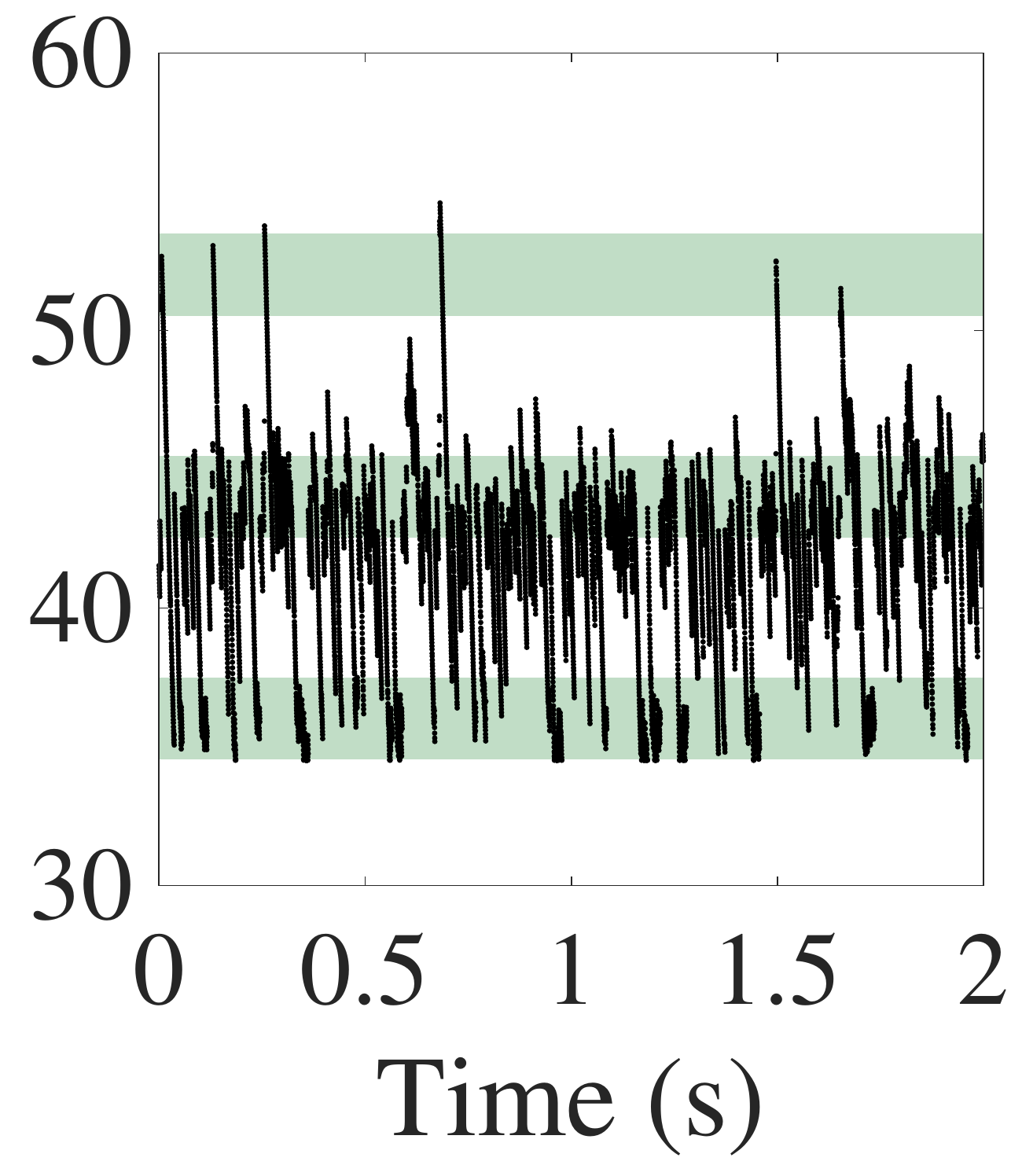}
        \caption{36~Mbit/s.}
        \label{fig:minDelay_c}
    \end{subfigure}
    \caption{Higher send rates (sub\hyp{}caption label) result in a higher probability of 
    transport block errors, so more packets encounter eight millisecond retransmission delays.}
    \label{fig:minDelay}
\end{figure}

The theoretical threshold, however, works poorly in practice because of the reordering operation.
We observe that the mobile user frequently buffers received 
packets in its reorder buffer (\S\ref{s:primer}), especially when 
offered load from the sender is high,  
causing significant fluctuations of packet delay. 
To demonstrate such an effect, we plot the measured one way delay 
at a mobile user under different sender offered loads in 
Figure~\ref{fig:minDelay}.  
We see that when offered load is low (6~Mbit/s), only a small portion of the received packets are 
retransmitted, as shown in Figure~\ref{fig:minDelay_a}. 
We also observe an approximate three millisecond network jitter introduced to the packet delay.  
When the offered load increases, the transport 
block error rate increases accordingly, as we have discussed 
in \S\ref{s:bandEst_Cellular}.  Consequently, the mobile user buffers more 
and more packets in its reorder buffer, introducing an multiple of 
eight ms retransmission delay to a increasing number of received packets, 
as shown in Figure~\ref{fig:minDelay_b} and~\ref{fig:minDelay_c}.
We note that, the minimum delay still captures the one way propagation delay, 
as there always are packets received correctly without retransmission and 
directly without buffering at the reorder buffer, \eg, the packets inside 
transport block of the first subframe in Figure~\ref{fig:reorder_buffer}.

According to the above analysis, we
set the switching threshold to $D_{\mathrm{th}}=\left(D_{\mathrm{prop}}+3\cdot8 +3\right)$~ms, 
where $\left(3\cdot8\right)$~ms accounts for the delay 
introduced by the three consecutive retransmissions 
(a transport block can be retransmitted at most three times~\cite{TS213}) 
and $3$~ms accounts for the network jitter (according 
to our experimental results, 94.1\% of the time, jitter is $\le 3$~ms).
To further mitigate the impact of greater network jitter 
and improve robustness, 
\systemname{} adds a threshold for the number of consecutive
packets with delay exceeding the delay threshold,
set to the number of packets $N_{\mathrm{pkt}}$ that can 
be transmitted over six subframes using current data rate:
\begin{equation}
N_{\mathrm{pkt}} = 6\cdot C_t / \mathrm{MSS}
\label{eqn:pack_thd}
\end{equation}
where $C_t$ is the current transport layer capacity with unit bits per 
subframe, and $\mathrm{MSS}$ is the maximum segment size.
We note that since our algorithm makes decisions based on relative delay,
\ie, the difference between current propagation delay and the threshold, 
instead of the absolute value of the delay,
\systemname{} does not require synchronization between the server and mobile clients.

\subsubsection{Internet Bottleneck State}\label{s:bandEst_Internet}

\begin{figure}[tp]
\centering
\includegraphics[width=0.95\linewidth]{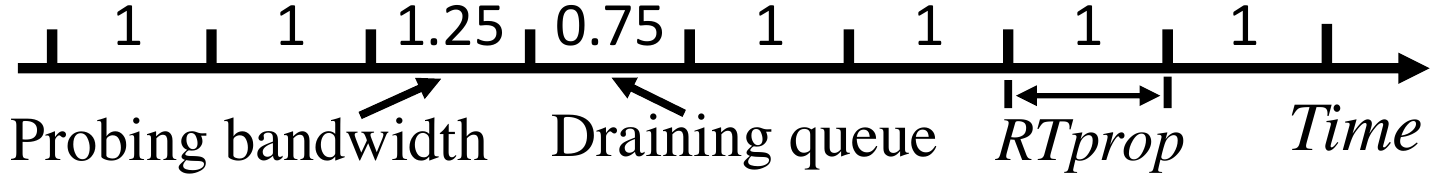}
\caption{BBR adopts a eight-phase cycle to probe the network 
bandwidth. The length of each phase is set to \emph{RTprop}.}
\label{fig:pacing_cycle}
\end{figure}

\systemname{} switches to a cellular\hyp{}tailored BBR to probe a rate that 
matches the capacity of the bottleneck link inside the Internet.
BBR senders estimate the bottleneck bandwidth of the connection 
($BtlBw$) as the maximum delivery rate in recent 10 RTTs,
and set their offered rate to $pacing\_gain\cdot BtlBw$. 
BBR's \emph{pacing\_gain} is set to 1.25 to probe possible 
idle bandwidth, to 0.75 when draining packets buffered in the 
previous probing period, and to one the rest of time. 
BBR's ProbeBW state repeats an eight\hyp{}phase cycle to probe bandwidth. 
The length of each phase is set to \emph{RTprop},
and the pacing gain in each phase is shown in 
Figure~\ref{fig:pacing_cycle}.
\systemname{} directly enters BBR's \textsf{ProbeBW} state,
then follows the same control logic as BBR to alternate between 
BBR's \textsf{ProbeBW}, \textsf{ProbeRTT}, \textsf{StartUp},
and \textsf{Drain} states. 

\parahead{Wireless\hyp{}aware, BBR\hyp{}like probing}
\systemname{} probes for a higher data rate that the 
Internet bottleneck supports, but also takes into account 
the fair\hyp{}share send rate of the cellular wireless link.
We adapt BBR's bandwidth probing scheme, changing the probing rate 
$C_{\mathrm{probe}}$ from a fixed $1.25 BtlBw$ to 
\begin{equation}
C_{\mathrm{probe}} = \min{\left\{1.25 BtlBw, C_{f}\right\}},
\label{eqn:c_probe}
\end{equation}
where $C_{f}$ is the maximum fair\hyp{}share capacity of the 
wireless link (estimated according to Eqn.~\ref{eqn:fair_rate} and translated 
to transport layer capacity according to Eqn.~\ref{eqn:cap} below).
The mobile user explicitly sends $C_{f}$ back to the sender when an
Internet bottleneck is detected. 
Similar to BBR, \systemname{} enters a draining phase after 
the probing phase to drain any buffered packets. 

When \systemname{} detects that the network is in the Internet\hyp{}bottleneck
state, there is already a packet queue formed inside the network. 
Therefore, before switching to handle that state, \systemname{} enters an additional 
draining phase that lasts for one \emph{RTprop}. 
During the draining phase, \systemname{} sets its
send rate to $0.5BtlBw$, leaving the remaining capacity of 
$0.5BtlBw$ for the bottleneck link to drain the packets buffered inside its queue. 

\parahead{Switching back to wireless bottleneck state}
If \systemnames{} send rate reaches $C_{f}$ without causing any packet queuing in the network, 
\ie, the mobile user observes $N_{\mathrm{pkt}}$ (calculated according to Eqn~\ref{eqn:pack_thd}) consecutive packets with delay smaller than $D_{\mathrm{th}}$~ms are observed at the mobile user, then
\systemname{} exits the Internet\hyp{}bottleneck state and 
re\hyp{}enters the wireless bottleneck state, staying in that state until the network is switched back to Internet\hyp{}bottleneck state.

\subsection{Fairness and TCP\hyp{}friendliness}

As it only modifies BBR's algorithms to be more conservative, \systemname{} is strictly
less aggressive than BBR when competing with flows sharing the same Internet bottleneck.
BBR's multi\hyp{}user fairness, RTT\hyp{}fairness and TCP\hyp{}friendliness have been 
well established in the literature\cite{BBR-eval1,BBR-eval2,BBR-eval3,BBR-fairness1}.

In the wireless bottleneck state,
multiple competing \systemname{} mobile clients quickly converge to a 
equilibrium with fair\hyp{}share cellular wireless capacity (as we demonstrate below
in \S\ref{s:fair_multi}), because each \systemname{} mobile client knows the 
number of competing users and their capacity usage in each aggregated cell
by decoding the cellular physical control channel, 
allowing it to explicitly calculate its fair\hyp{}share capacity 
(\S\ref{s:fair_share}) and then guide its sender to match its
sending rate accordingly.
In contrast, conventional end\hyp{}to\hyp{}end congestion control 
algorithms need to probe the fair\hyp{}share of bottleneck 
capacity with a more complicated series of probing and backoff steps, 
which is less efficient.
\systemname{} also fairly shares wireless link capacity 
with existing congestion control algorithms, \eg, CUBIC and BBR,
with the help of cell tower's fairness policy, as our experimental 
evaluation later demonstrates (\S\ref{s:fair_TCP}).

\systemname{} flows with different propagation delays fairly share 
wireless capacity (as we demonstrate in \S\ref{s:fair_RTT}), 
because of two reasons, one from the design of \systemname{} and 
one from the buffer structure of base station. 
First, \systemname{} explicitly calculates the fair-share capacity,
while most conventional congestion control algorithm adopt 
additive\hyp{}increase multiplicative\hyp{}decrease (AMID) schemes 
to probe for the fair share. 
During the additive increase, the sender of a flow with smaller 
propagation delay increases its window faster than flows with 
larger delay, resulting in unfairness~\cite{CUBIC,RTTfair}. 
Second, the base station provides separate buffers for every user, which prevents 
large\hyp{}\emph{RTprop} connections from dominating the bottleneck buffer. 
For example, a BBR connection with a large \emph{RTprop} calculates a large 
BDP and thus injects significant amount of inflight packets into the network,
which queue at the bottleneck buffer and lower the delivery rate for another 
BBR flow with a small \emph{RTprop} and hence a small number of inflight packets. 
The separate buffer at cellular base station isolates the inflight packets from different 
flows sharing the wireless link and thus prevents unfairness.

\section{Implementation}
\label{s:impl}

\begin{figure}[tbh]
    \centering
    \begin{subfigure}[b]{0.58\linewidth}
        \centering
        \includegraphics[width=0.99\textwidth]{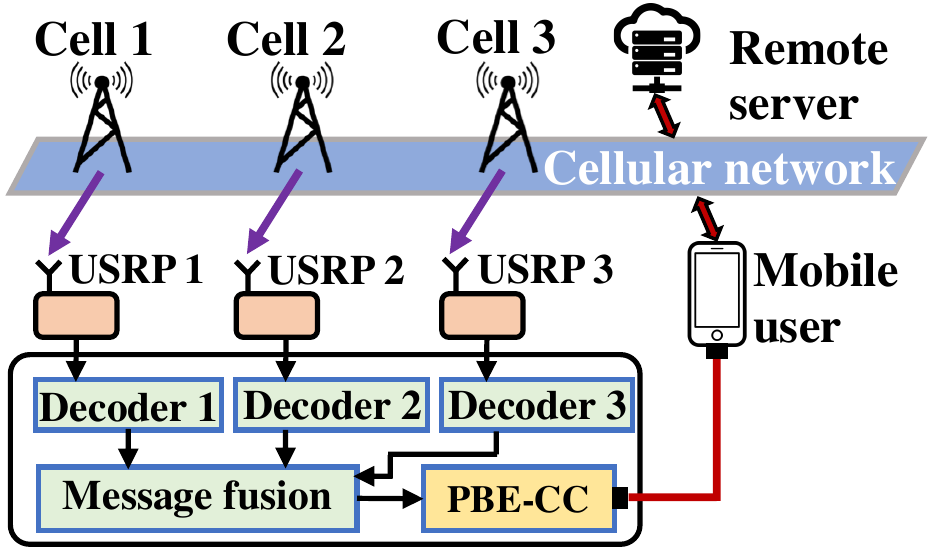}
        \caption{Prototyping platform.}
        \label{fig:imp_platform}
    \end{subfigure}
    \hfill
    \begin{subfigure}[b]{0.41\linewidth}
        \centering
        \includegraphics[width=0.99\textwidth]{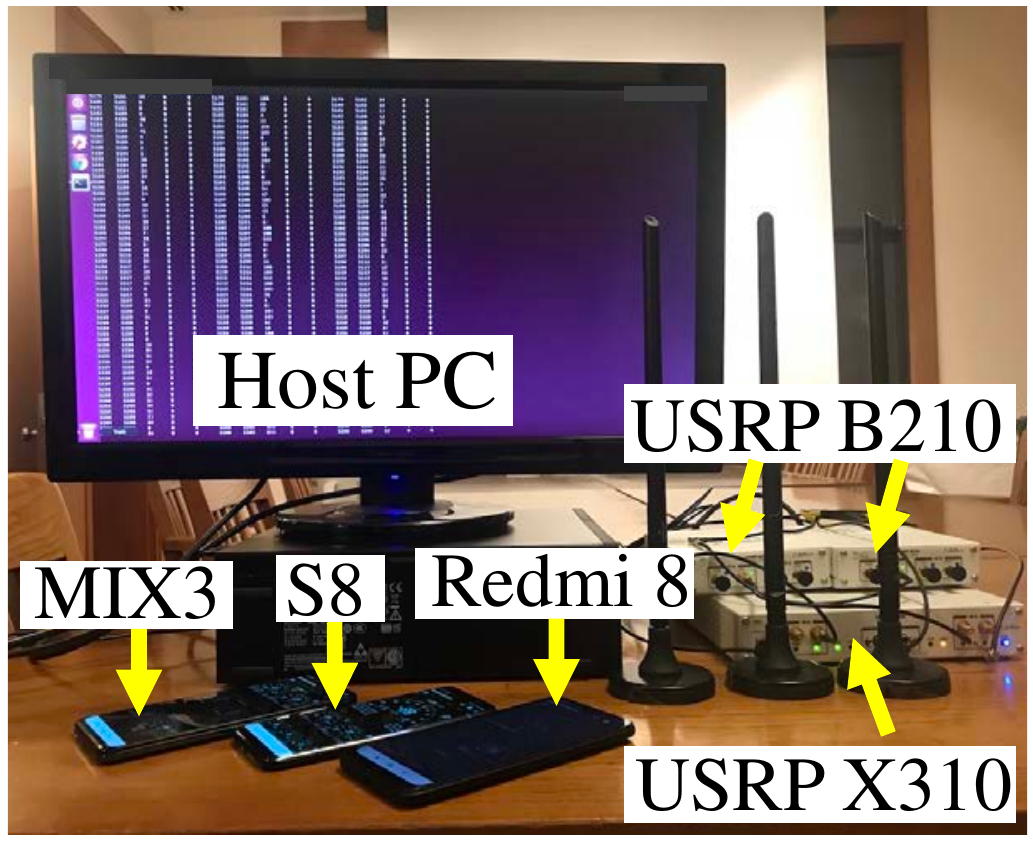}
        \caption{System setup.}
        \label{fig:imp_testbed}
    \end{subfigure}
    \caption{The architecture of the open\hyp{}source
    \systemname{} cellular congestion control prototyping platform \textbf{(a)}.
    The setup of \systemname{} mobile clients is shown in \textbf{(b)}.}
    \label{fig:imp_system}
\end{figure}

Programming a mobile phone to decode every control message transmitted over 
the control channel requires customization of the cellular 
firmware inside the phone.  The source code of current cellular firmware, however, 
is proprietary to cellular equipment manufacturers, thus is not accessible.
As a proof of concept, we build an open\hyp{}source congestion control 
prototyping platform that supports control message decoding, 
bypassing the need to customize firmware.      
The key component of our platform is an open\hyp{}source control channel 
decoder that uses an off\hyp{}the\hyp{}shelf software defined radio (USRP in 
our implementation) as the RF front\hyp{}end to collect cellular wireless signals, 
and a PC as the host to decode the control messages from the collected signals.
We start multiple parallel control channel decoders, 
each decoding the signal from one cell in the list of aggregated cells of the 
mobile user, as shown in Figure~\ref{fig:imp_platform}.
Our \textsf{Message Fusion} module aligns the decoded control messages from 
multiple decoders according to their subframe indices, 
feeding the aligned messages to our \textsf{Congestion Control} module.

We implement our cellular control channel decoder in 3,300 lines of C code 
(excluding reused code).  We reuse the physical layer signal processing modules from 
an open\hyp{}source LTE library (\emph{srsLTE}~\cite{srslte}), \ie, a wireless 
channel estimator, a demodulator, and a convolutional decoder.
Each decoder decodes the control channel by searching every possible 
message position inside the control channel of one subframe and trying
all possible formats at each location until finding the correct message.\footnote{The 
3GPP standard defines 10 formats for control messages~\cite{TS212}. 
The base station does not explicitly indicate the format of the message it sends.}
We implement the parallel decoding structure using multi\hyp{}threading, 
allowing one PC to decode the control channel of multiple cells simultaneously.
In our test, a six\hyp{}core PC is able to decode six cell towers while maintaining
CPU usage of each core below 40 percent.
We will open\hyp{}source our platform to facilitate future cross\hyp{}layer 
cellular congestion control design and prototyping. 

We implement a user\hyp{}space, UDP\hyp{}based prototype of 
\systemnames{} congestion control algorithm using 874 lines of C++ code 
(517 on the mobile client side and 357 at the sender side).
The client\hyp{}side \systemname{} module takes the decoded control messages 
as input, and communicates with the sender side via a commercial mobile 
phone tethered with the host PC, as shown in Figure~\ref{fig:imp_platform}.
When the \systemname{} mobile client receives a data packet, it estimates the one
way packet propagation delay $D_{\mathrm{prop}}$ (\S\ref{S:track_alternation}),
and feeds back the estimated capacity. 
We describe the capacity using an interval in milliseconds between sending 
two 1500\hyp{}byte packets, and represent it with a 32\hyp{}bit integer. 
The \systemname{} client also identifies the current bottleneck state, notifying
the sender via one bit in the ACK.  When the \systemname{} sender receives an
ACK, it sets its sending rate to the capacity indicated therein.
The \systemname{} sender also updates its estimated \emph{RTprop} and 
\emph{BtlBw} with every received ACK, so it can immediately switch to the 
cellular\hyp{}tailored BBR if and when the bottleneck location changes.
\section{Evaluation}\label{s:eval}

In this section, we evaluate the performance of \systemname{} in a commercial cellular network and compare with existing end\hyp{}to\hyp{}end congestion control algorithms. 

\subsection{Methodology}
\parahead{Content senders}
We configure Amazon AWS servers as the \systemname{} senders. 
To evaluate \systemnames{} performance over flows with significantly different RTT, 
we setup AWS servers at different continents, \ie, three in US and one in Singapore.

\parahead{Mobile clients}
Each \systemname{} mobile client is a combination of multiple USRPs for 
signal collection, a host PC for control channel decoding,
and a commercial mobile phone for cellular communication,
as shown in Figure~\ref{fig:imp_testbed}.
We use both USRP X310~\cite{X310} and B210~\cite{B210} in our implementation. 
The host PC we use for each mobile client is a Dell OptiPlex 7060 
(Intel Core i7\hyp{}8700 CPU, 16~GB RAM, and Ubuntu 16.04).
We use various types of mobile phones that support carrier aggregation in hardware, 
including a Xiaomi MIX3, a Redmi~8, and a Samsung~S8.
The cellular network configures the same primary cell for all three 
phones, but different numbers of aggregated cells for each phone, 
\ie, only one cell for the Redmi 8, two cells for the MIX3 and three cells for the S8.

\parahead{Congestion control algorithms to compare}
We compare \systemname{} against seven end\hyp{}to\hyp{}end congestion control algorithms, 
including algorithms specially designed for cellular networks 
like Sprout~\cite{Sprout} and Verus~\cite{Verus},
algorithms that have already been included inside the official Linux 
kernel like BBR~\cite{BBR} and CUBIC~\cite{CUBIC},
and recently\hyp{}proposed algorithms like Copa~\cite{Copa}, 
PCC~\cite{PCC} and PCC\hyp{}Vivace~\cite{PCC-v}. 
We test all the above algorithms in commercial cellular 
networks covering our campus using Pantheon~\cite{Pantheon}.

\begin{figure}[tbh]
    \centering
    \begin{subfigure}[b]{0.49\linewidth}
        \centering
        \includegraphics[width=0.99\textwidth]{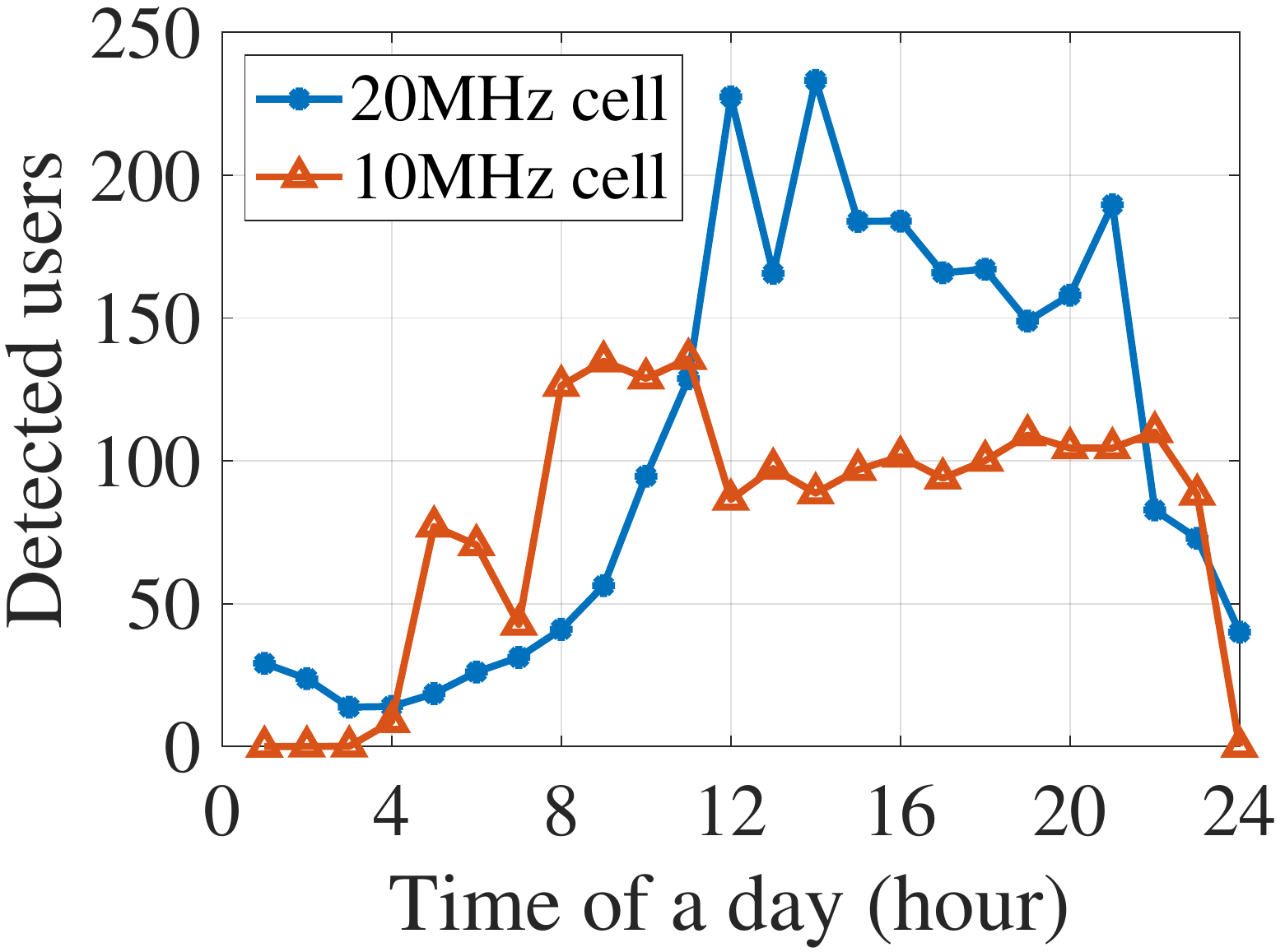}
        \caption{Detected users. }
        \label{fig:cellStat_users}
    \end{subfigure}
    \hfill
    \begin{subfigure}[b]{0.49\linewidth}
        \centering
        \includegraphics[width=0.99\textwidth]{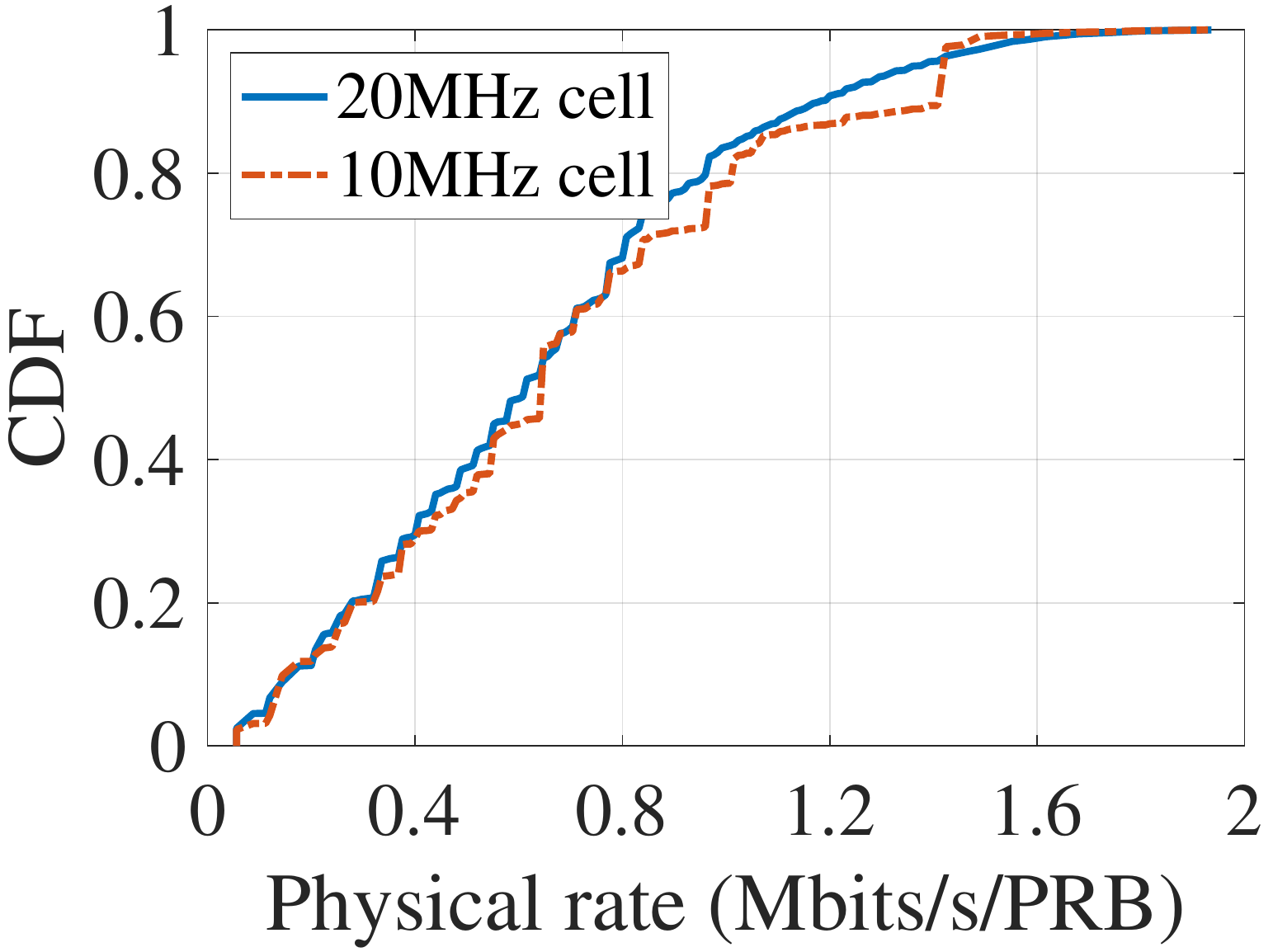}
        \caption{Physical data rate. }
        \label{fig:cellStat_phyRate}
    \end{subfigure}
    \caption{\textbf{(a)} The number of detected users in each hour of a day that have data communication with two base stations ( a 20~MHz one and a 10~MHz one). 
        \textbf{(b)} The distribution of wireless physical data rate of the detected users. 
    }
    \label{fig:cellStatus}
    \vspace{-0.4cm}
\end{figure}

\subsection{Micro-benchmark: Cell Status}
In this section, we perform a micro-benchmark to present two important statistics of the cell tower: 
(1) the number of users that have communicated with the cell tower in each hour and (2) the distribution of wireless physical data rate of the users. 
We leverage our control channel decoder to decode the control messages that two base stations (a 20~MHz one and a 10~MHz one) transmit. 
We conduct the experiments for 24 hours and count the number of active users in each hour. 
We plot the result in Figure~\ref{fig:cellStat_users}, 
from which we see that each cell serves a large number of users during peak hours of a day, 
\eg, during the 12 to 20 hours period, the average number of users per hour is 181 and 97 for 20~MHz and 10~MHz cell, respectively. 
Furthermore, the number of users varies significantly within a day, 
\ie, maximum 233 and 135 users, minimum 13 and zero users for 20~MHz and 10~MHz cell, respectively. 
We note that the 10~MHz cell is turned off by the operator during zero to three hour period, so we observe zero users.
We also plot the distribution of the wireless physical data rate of all detected users, in Figure~\ref{fig:cellStat_phyRate}.
We see that even though the users has diverse data rates, a large portion are low-rate users,
\eg, 77.4\% and 71.9\% users have rate smaller than half of the maximum achievable data rate (1.8~Mbit/s/PRB),
for 10~MHz and 20~MHz cell, respectively.
In the following sections, we evaluate the performance of \systemname{} 
working atop of these cells that serve large number of diverse users. 

\subsection{End\hyp{}to\hyp{}end Delay and Throughput}\label{s:delay_thput}
In this section, we investigate the delay and throughput 
performance of \systemname{} achieved in a commercial cellular network.

\subsubsection{Performance of Stationary Cellular Links}
We investigate \systemnames{} performance on stationary cellular links. 
We build connections between servers and stationary mobile users 
over which senders transmit to their corresponding users for 20 seconds, 
recording achieved throughput, packet delay, and arrival time in each flow.
We change the congestion control algorithm the sender adopts and test eight algorithms sequentially. 
Since the capacity of the cellular network varies when testing each algorithm, 
we repeat the whole preceding test sequence (sequentially testing all algorithms) five times at one location to provide a fair comparison of achieved throughput, across different congestion control algorithms.
Furthermore, we conduct the foregoing experiment using different phones, in 
order to measure performance with different numbers of aggregated cells.
We repeat these experiments at multiple indoor and outdoor locations and 
at different times of the day, \ie, 
daytime when the cell is busy, and late night when the cell is idle. In total, 
we test 40 locations, 
covering all combinations of indoor\fshyp{}outdoor, 
one\fshyp{}two\fshyp{}three aggregated cells and busy\fshyp{}idle links.

\begin{figure}[t]
    \centering
    \begin{subfigure}[b]{0.49\linewidth}
        \centering
        \includegraphics[width=0.99\textwidth]{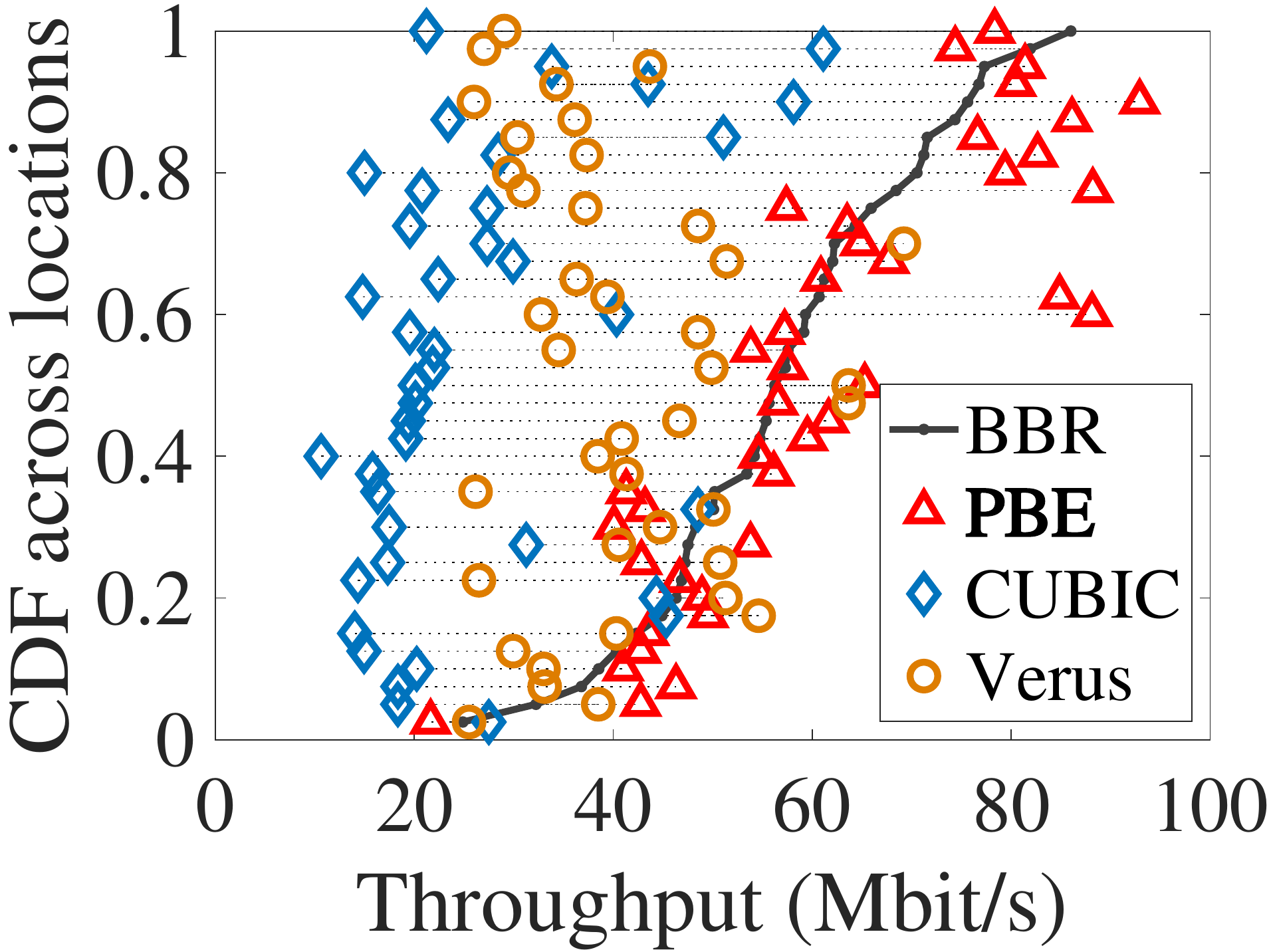}
        \caption{Average throughput.}
        \label{fig:static_thput_cdf}
    \end{subfigure}
    \hfill
    \begin{subfigure}[b]{0.49\linewidth}
        \centering
        \includegraphics[width=0.99\textwidth]{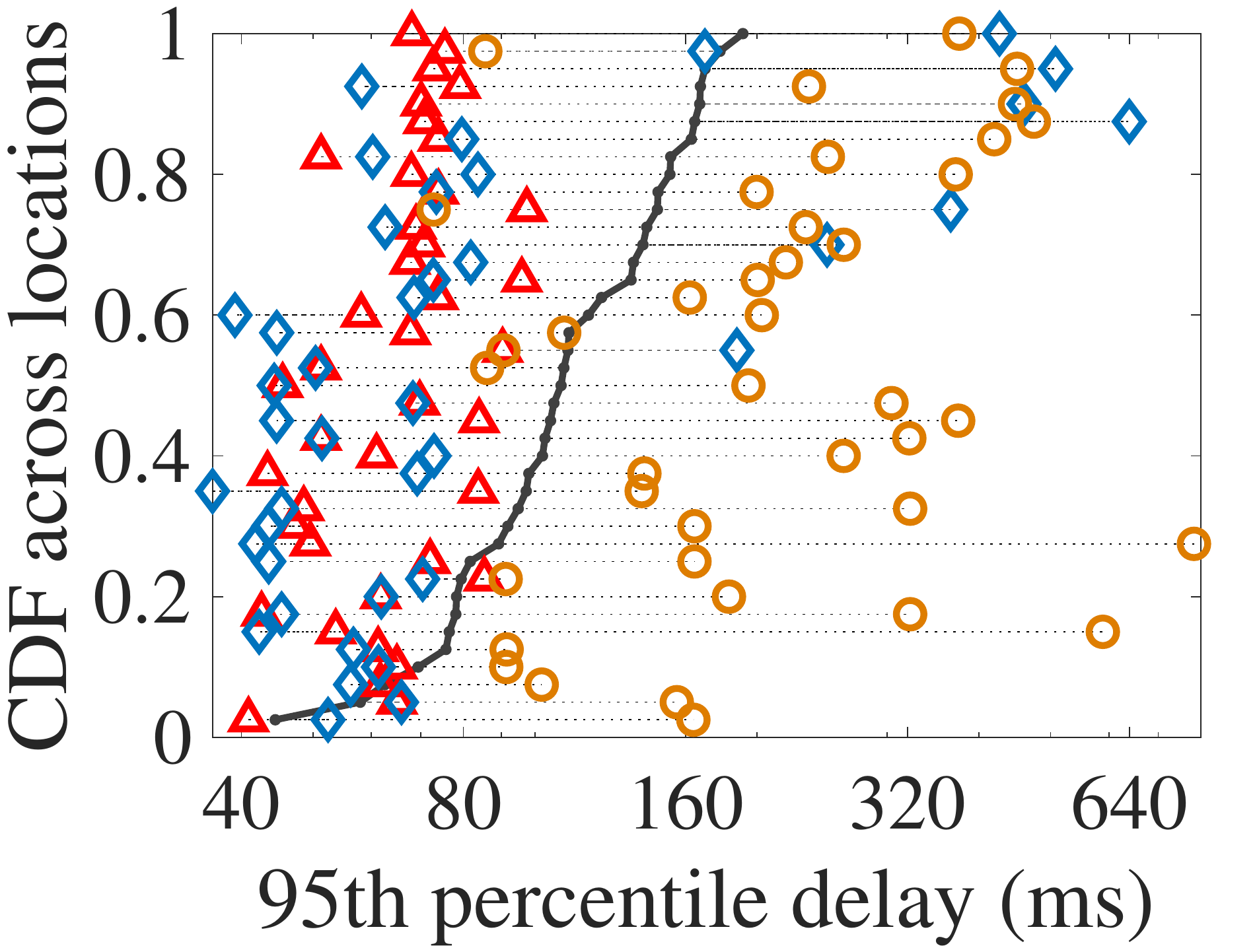}
        \caption{$95^{\mathrm{th}}$ percentile delay.}
        \label{fig:static_delay_cdf}
    \end{subfigure}
    \caption{The distribution of throughput~(\textbf{a}) and $95^{\mathrm{th}}$ 
    percentile delay~(\textbf{b}), of \systemname{}, BBR, Verus, and 
    CUBIC (the four \emph{``high throughput''} algorithms), across 40 locations.}
    \label{fig:delay_thput_40loc}
    \vspace{-0.4cm}
\end{figure}

\begin{figure*}[tbh]
    \centering
    \begin{subfigure}[b]{0.24\linewidth}
        \centering
        \includegraphics[width=0.99\textwidth]{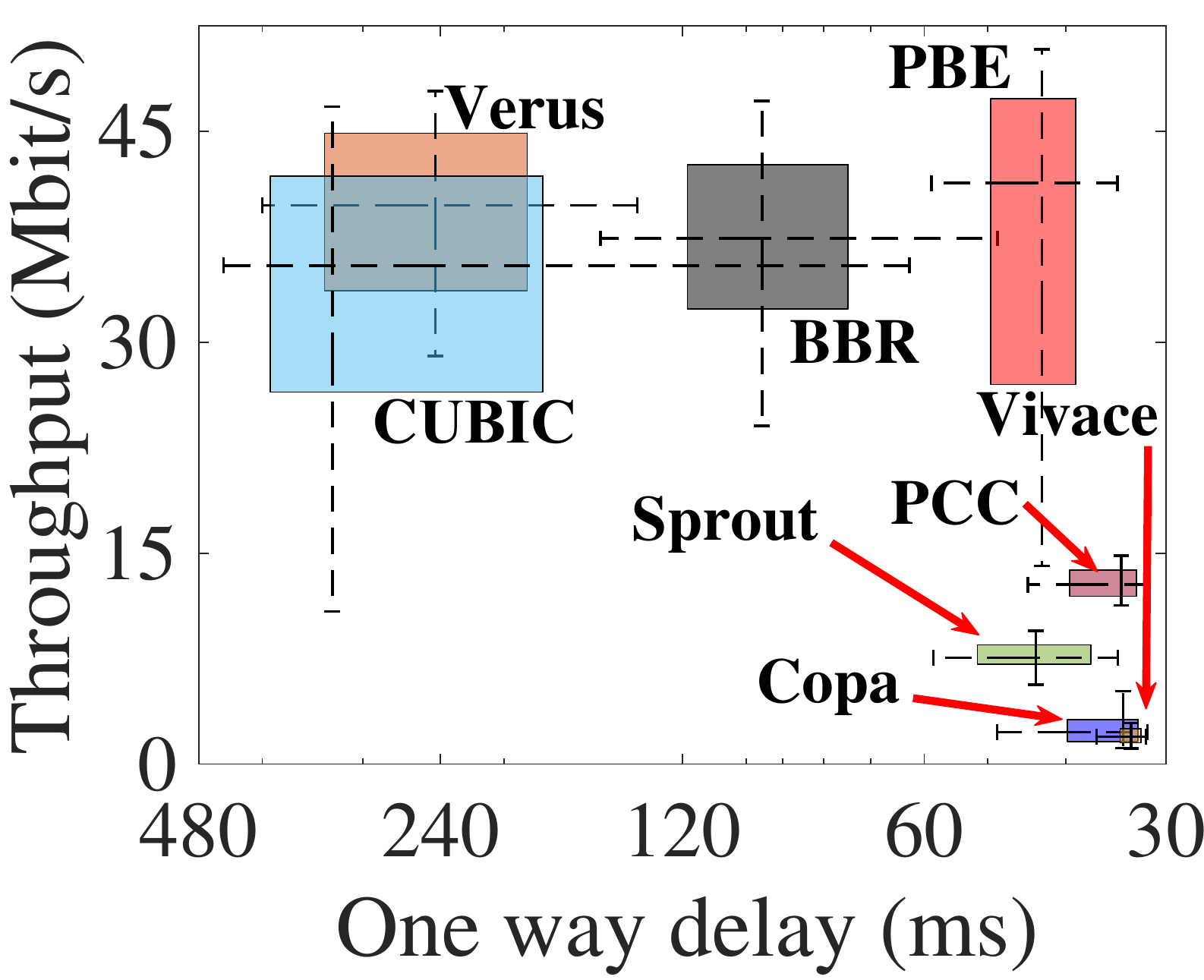}
        \caption{One aggregated cell, indoor and busy hours.}
        \label{fig:dt_1cell_in_busy}
    \end{subfigure}
    \hfill
    \begin{subfigure}[b]{0.24\linewidth}
        \centering
        \includegraphics[width=0.99\textwidth]{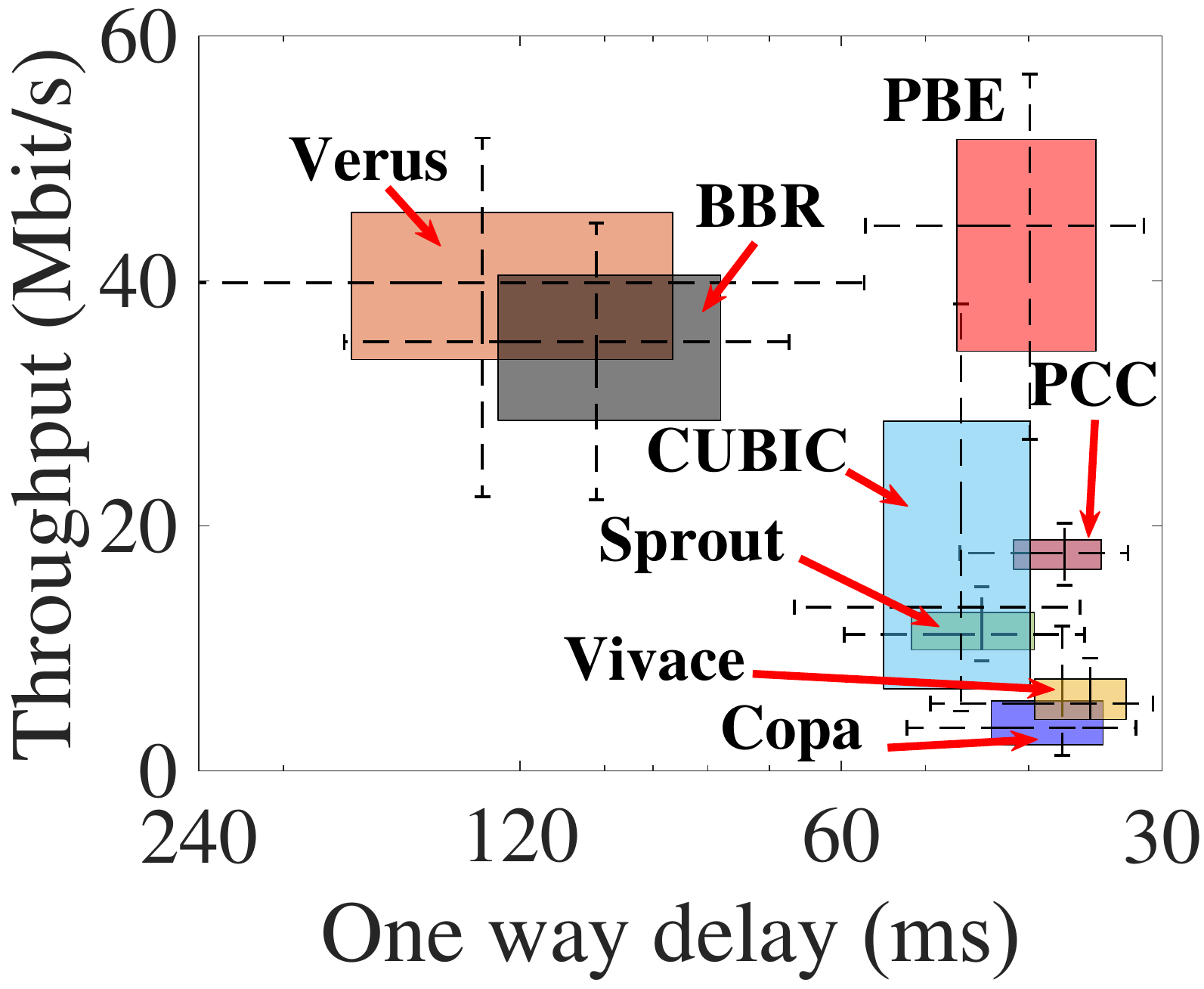}
        \caption{Two aggregated cells, indoor and busy hours.}
        \label{fig:dt_2cell_in_busy}
    \end{subfigure}
    \hfill
    \begin{subfigure}[b]{0.24\linewidth}
        \centering
        \includegraphics[width=0.99\textwidth]{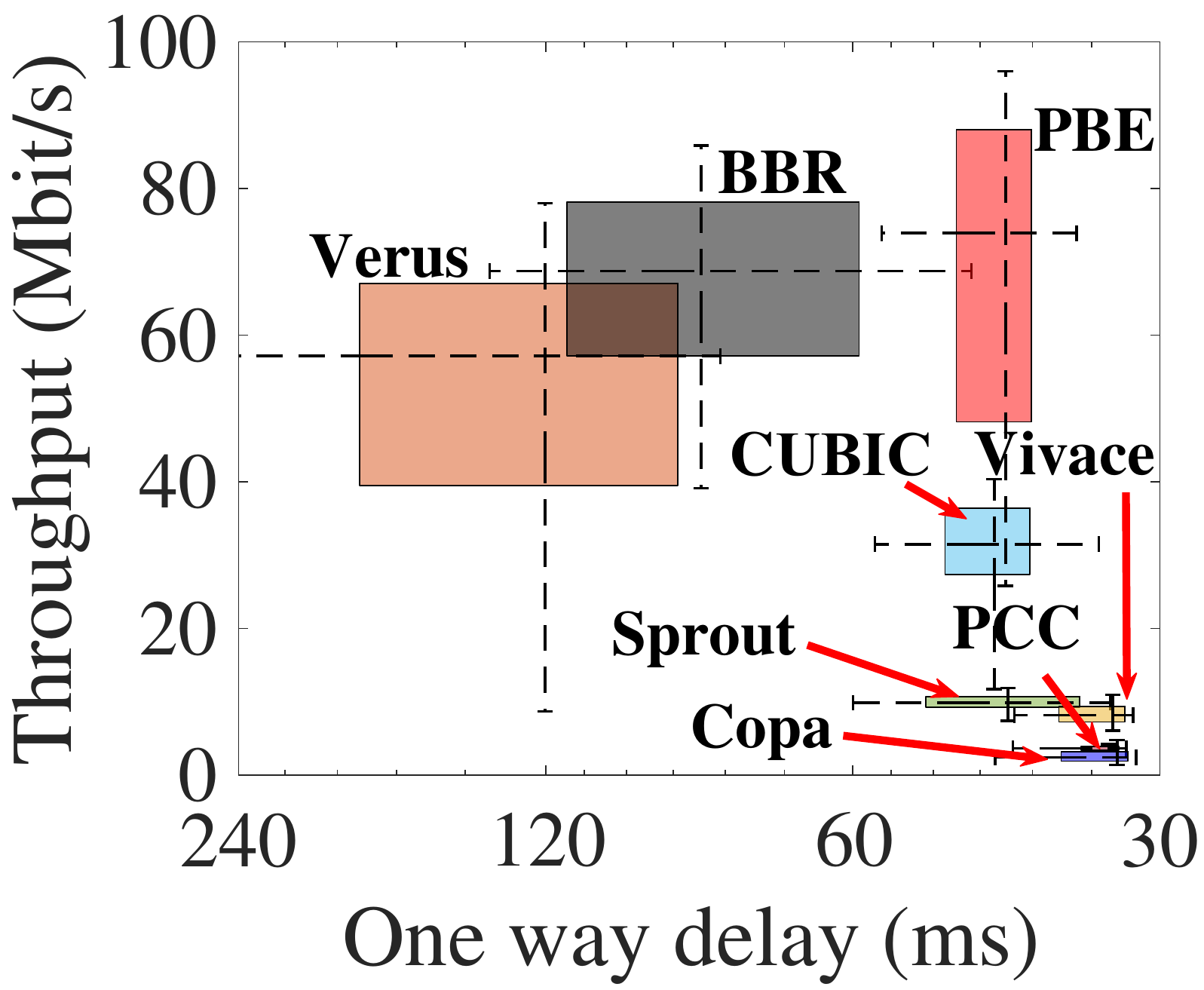}
        \caption{Three aggregated cells, indoor and busy hours.}
        \label{fig:dt_3cell_in_busy}
    \end{subfigure}
    \hfill
    \begin{subfigure}[b]{0.24\linewidth}
        \centering
        \includegraphics[width=0.99\textwidth]{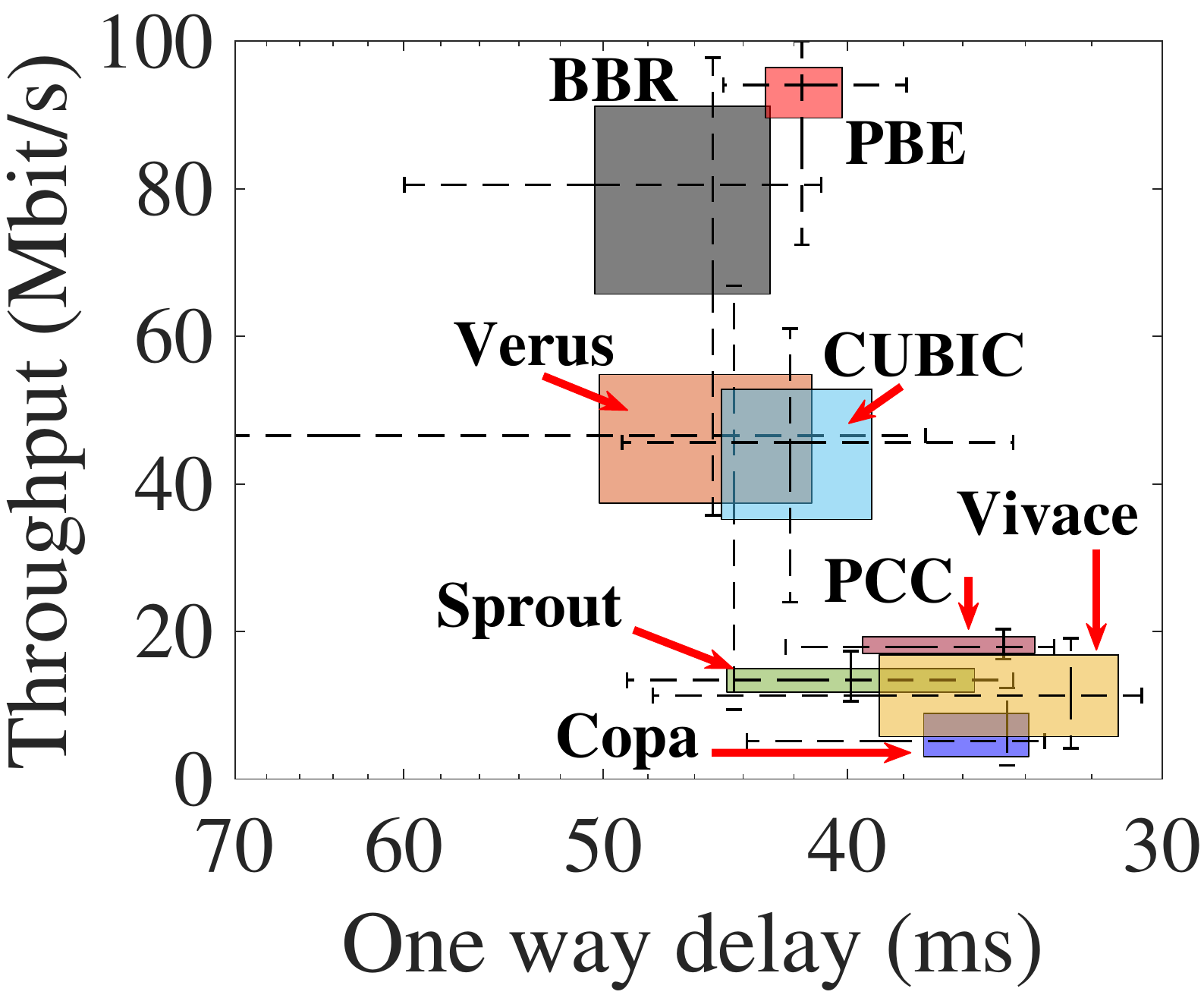}
        \caption{Three aggregated cells, indoor and late night (idle cells).}
        \label{fig:dt_3cell_in_idle}
    \end{subfigure}
    \caption{One way packet delay and throughput achieved by eight congestion control algorithms. 
    The right and lower edge of the box represents the 25\% percentile of the achieved delay and throughput, respectively.
    The left and upper edge give the 75th percentiles. 
    The two ends of the error bar gives the 10th and 90th percentiles. 
    The intersection point of the horizontal and vertical error bar represents the median of achieved delay and throughput.}
    \label{fig:delay_thput}
    \vspace{-0.3cm}
\end{figure*}

\parahead{Comparison among high-throughput algorithms}
As we will demonstrate in the following section, 
\systemname{}, BBR, CUBIC, and Verus achieve significantly higher throughput than the
other four algorithms we examine.
We plot the distribution of the averaged throughput and $95^{\mathrm{th}}$
percentile one way delay achieved by these four algorithms, 
in Figure~\ref{fig:static_thput_cdf} and~\ref{fig:static_delay_cdf}.
We see that \systemname{} achieves the highest throughput for most of the 
stationary links, while simultaneously maintaining very low latency. 
Table~\ref{t:summary_perf} on p.~\pageref{t:summary_perf} 
summarizes the performance improvement of \systemname{} over BBR and Verus. 
\systemname{} achieves $2.3\times$ average higher throughput than CUBIC,
and at the same time reduces $95^{\mathrm{th}}$ percentile delay by 
$1.8\times$.

\begin{figure}
    \centering
        \begin{subfigure}[b]{0.49\linewidth}
            \centering
            \includegraphics[width=0.99\textwidth]{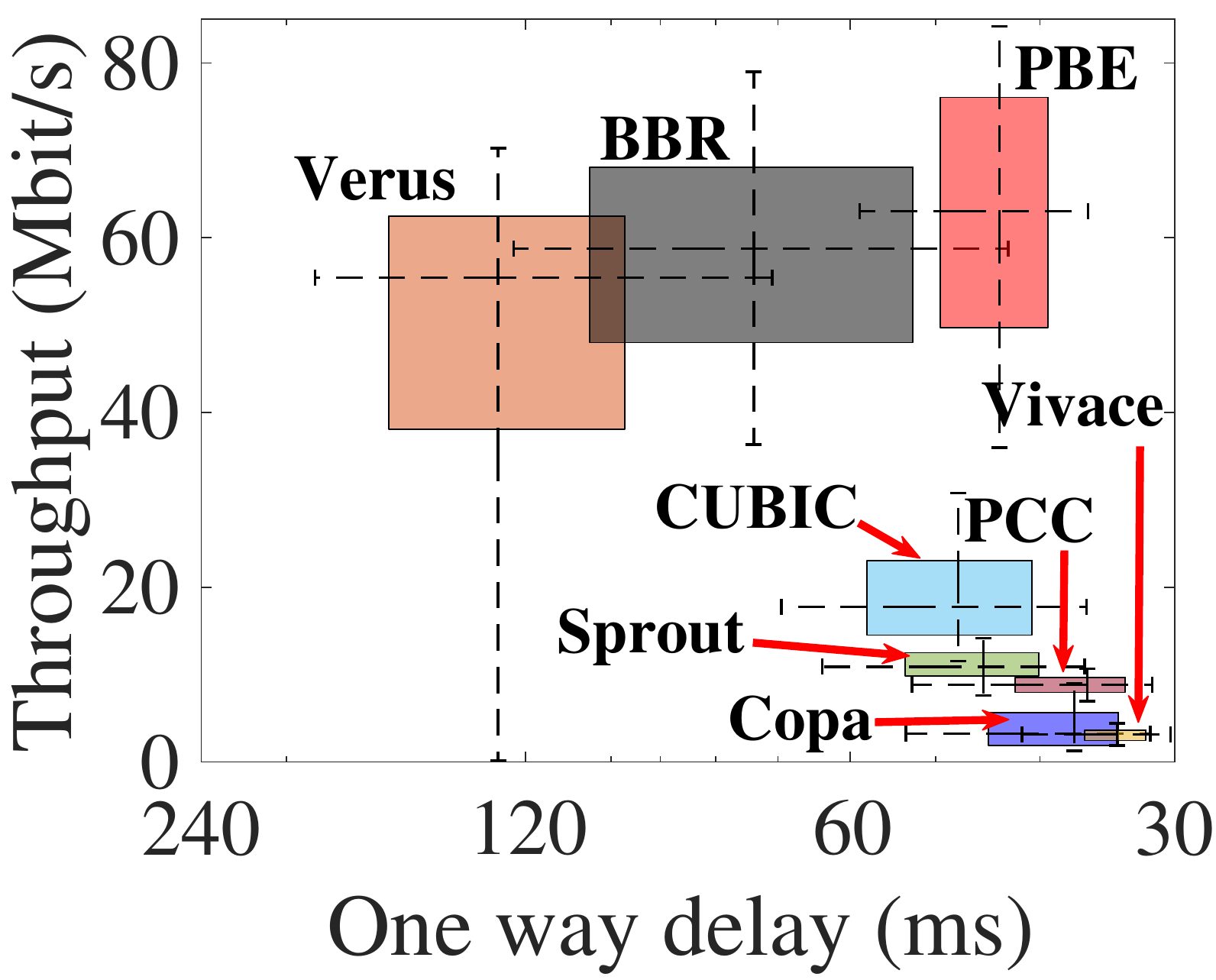}
            \caption{Two aggregated cells, outdoor and busy hours.}
            \label{fig:dt_2cell_out_busy}
        \end{subfigure}
        \hfill
        \begin{subfigure}[b]{0.49\linewidth}
            \centering
            \includegraphics[width=0.99\textwidth]{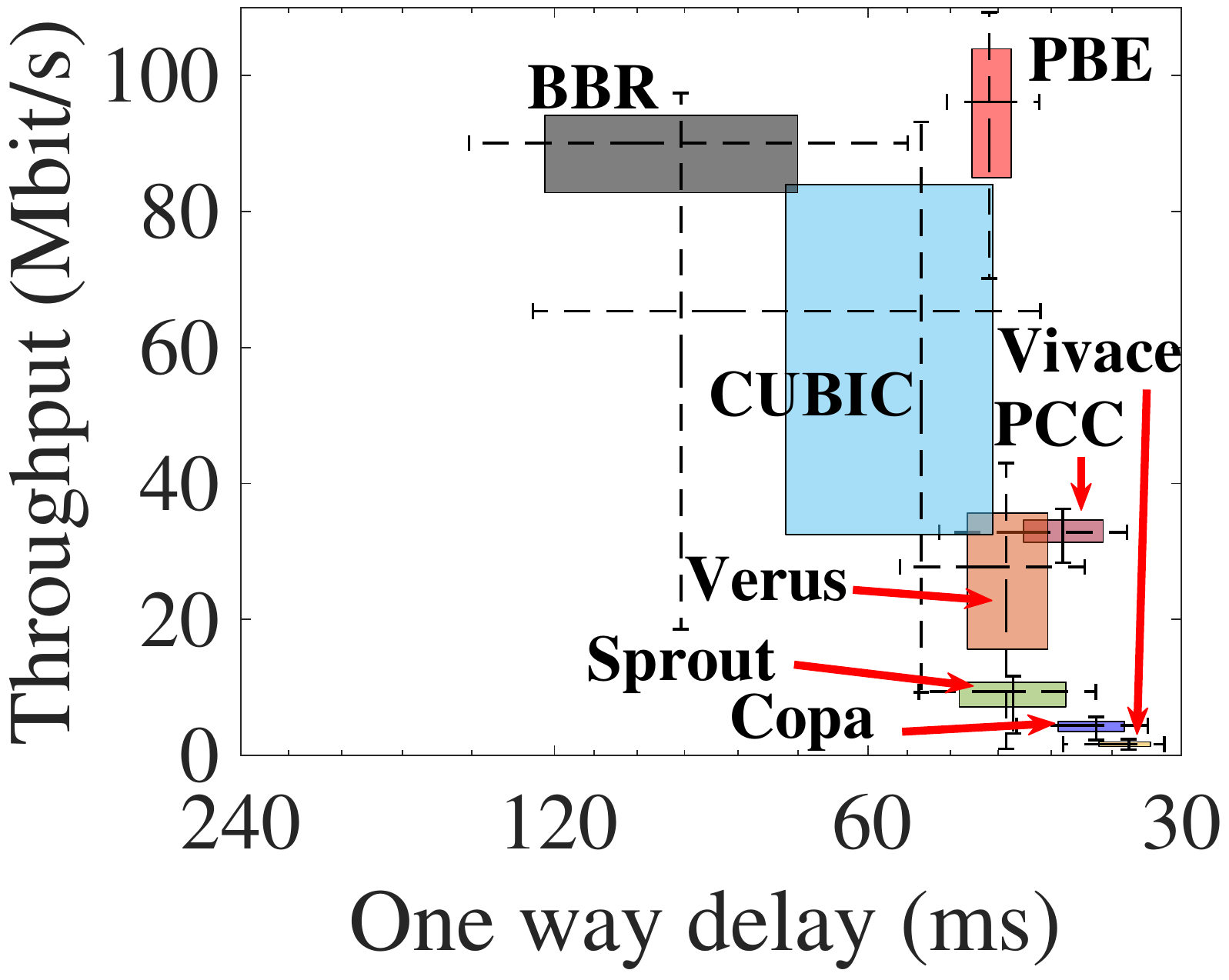}
            \caption{Two aggregated cells, outdoor and late night (idle).}
            \label{fig:dt_2cell_out_idle}
        \end{subfigure}
        \caption{The oneway packet delay and throughput achieved by eight congestion control algorithms in two different outdoor tests covering the busy and idle cell status.}
        \label{fig:delay_thput_2cell_inout}
\end{figure}

\begin{figure}
    \begin{minipage}[htb]{0.49\linewidth}
        \centering
        \includegraphics[width=0.99\linewidth]{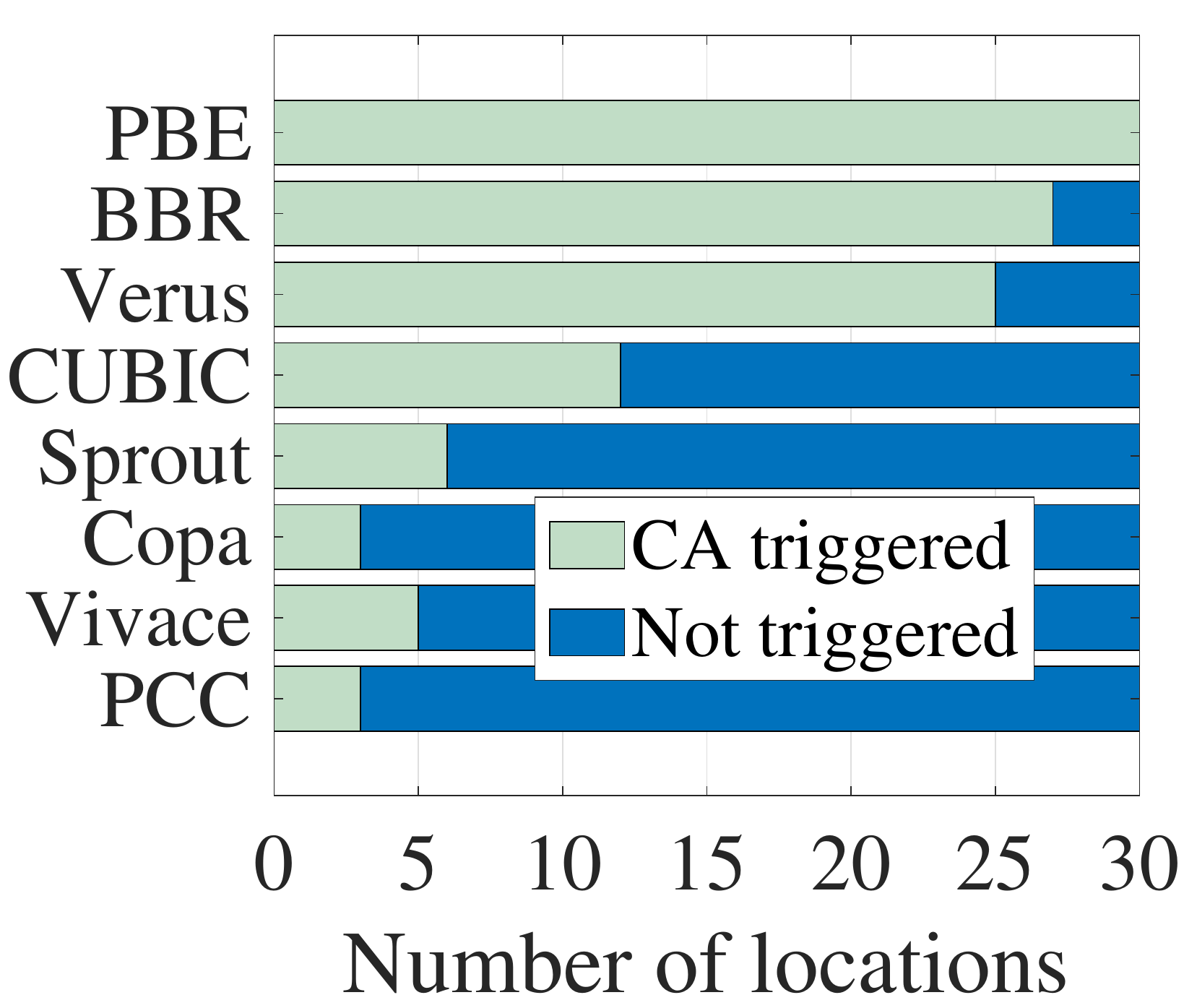}
        \caption{The number of locations at which each congestion control algorithm triggers carrier aggregation.}
        \label{fig:ca_triggered}
    \end{minipage}
    \hfill
    \begin{minipage}[htb]{0.47\linewidth}
         \centering
        \includegraphics[width=0.99\linewidth]{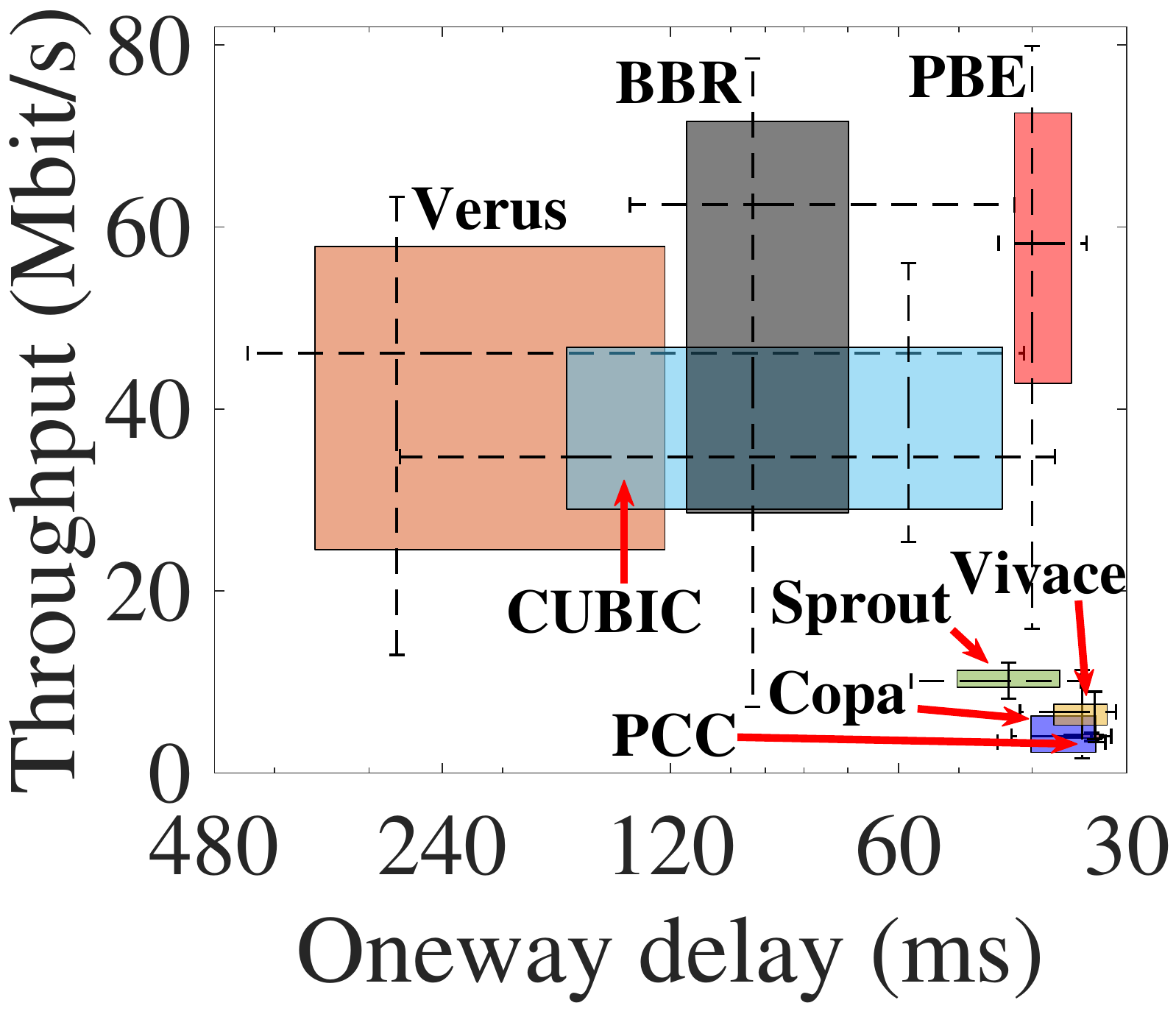}
        \caption{The achieved delay and throughput of when the mobile client is moving along the same trajectory.}
        \label{fig:mobility_delay_thput}
    \end{minipage}
\end{figure}

\parahead{Detailed comparison among eight algorithms}
To provide a detailed performance comparison among all eight algorithms, 
we select six representative locations,
and plot the $10^{\mathrm{th}}$, $25^{\mathrm{th}}$, $50^{\mathrm{th}}$, $75^{\mathrm{th}}$,
and $90^{\mathrm{th}}$ percentile throughputs (averaged 
over every 100\hyp{}millisecond interval) and delay, for eight algorithms,
in Figures~\ref{fig:delay_thput} and~\ref{fig:delay_thput_2cell_inout}. 
We have three observations from these figures.
First, \systemname{} achieves high average throughput, but also has somewhat
high throughput variance, since \systemname{} is able to match its send 
rate to the varying wireless channel capacity.
BBR achieves comparable throughput with \systemname{} in all selected locations,
but with higher delay.
Verus, a congestion control algorithm designed for 
cellular networks, also achieves relatively high throughput in many 
locations, but introduces excessive packet delays.
The performance of CUBIC is highly unpredictable, alternating between 
high throughput (but high delay) and low throughput (but low delay), as
our order statistics demonstrate.
The other four algorithms, including Copa, PCC, PCC\hyp{}Vivace, 
and Sprout, have a large throughput disadvantage compared to \systemname{}.
We plot the number of locations at which each congestion control algorithm 
triggers the cellular network to activate secondary cells for providing extra
throughput (maximum 30 locations, since we use Redmi~8 that uses only one cell, in 10 locations), 
in Figure~\ref{fig:ca_triggered}.
We see that Copa, PCC, PCC\hyp{}Vivace, and Sprout use very 
conservative send rates, so the cellular network disables
carrier aggregation at most locations, resulting in significant 
under\hyp{}utilization of the available wireless capacity. 

\begin{figure}[htp]
    \centering
    \includegraphics[width=0.99\linewidth]{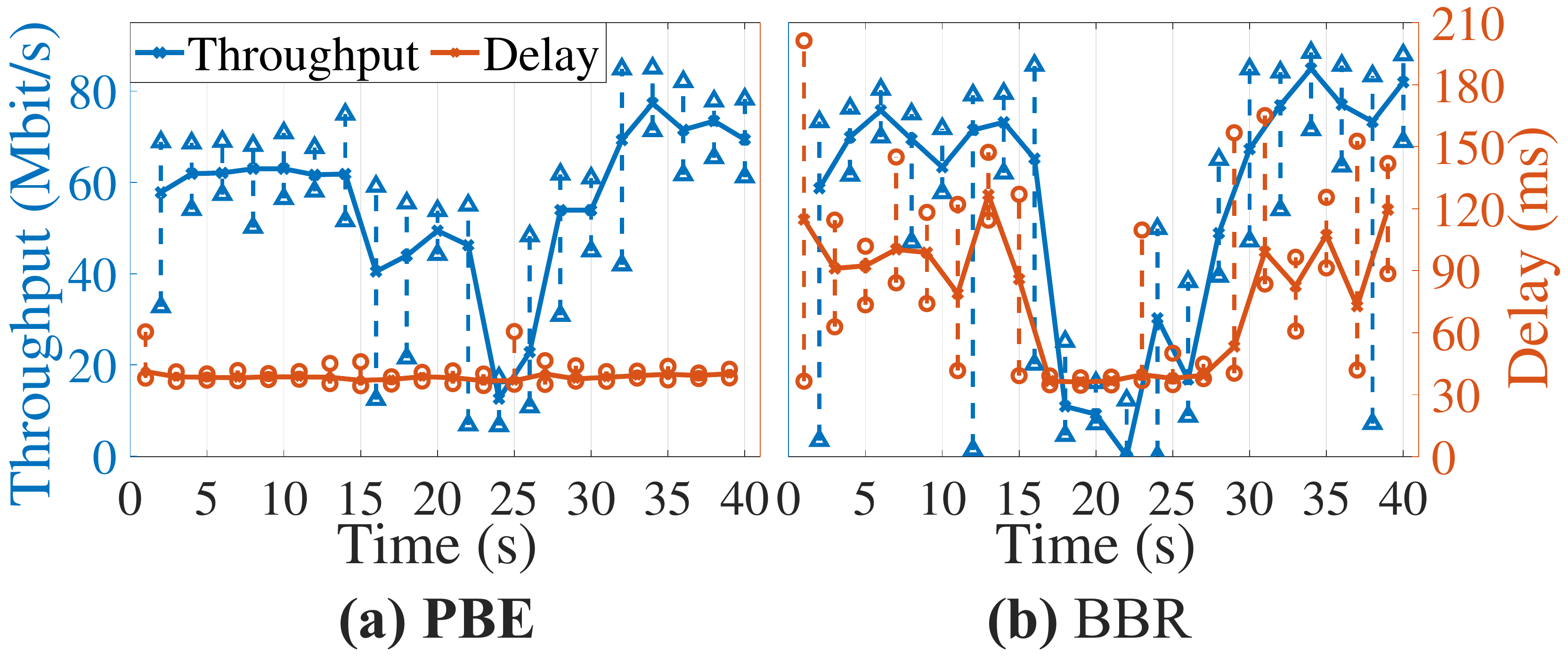}
    \caption{Delay and throughput achieved by \systemname{}~(\textbf{a}) and BBR~(\textbf{b})
    when the user is moving along the same trajectory.}
    \label{fig:mobility_BBR_PBE}
    \vspace{-0.4cm}
\end{figure}

\systemname{} achieves a low delay and delay variance. 
Comparing against BBR and Verus, two algorithms with relatively high throughput, 
\systemname{} incurs much smaller delays. However, \systemname{} 
has a slightly higher latency than the four algorithms with low throughput.
Such a delay gap is mainly caused by cellular retransmissions:
as we have demonstrated in Figure~\ref{fig:TB_err_TB_size}, 
higher throughputs result in a larger TB error rates,
and thus more retransmissions.  Therefore, under schemes with higher
throughput, slightly more packets incur
a multiple of eight millisecond retransmission delay. 

Finally, we observe that \systemname{} has low variance in both delay and 
throughput when cells are idle, as shown in Figures~\ref{fig:dt_3cell_in_idle} and~\ref{fig:dt_2cell_out_idle}.
Without competing traffic and mobility, wireless capacity 
becomes stable for a static user in an idle cell.
\systemname{} then achieves stable throughput and delay by accurately 
estimating this capacity.
\begin{figure*}[htb]
     \begin{minipage}[htb]{0.24\linewidth}
        \centering
        \includegraphics[width=0.98\textwidth]{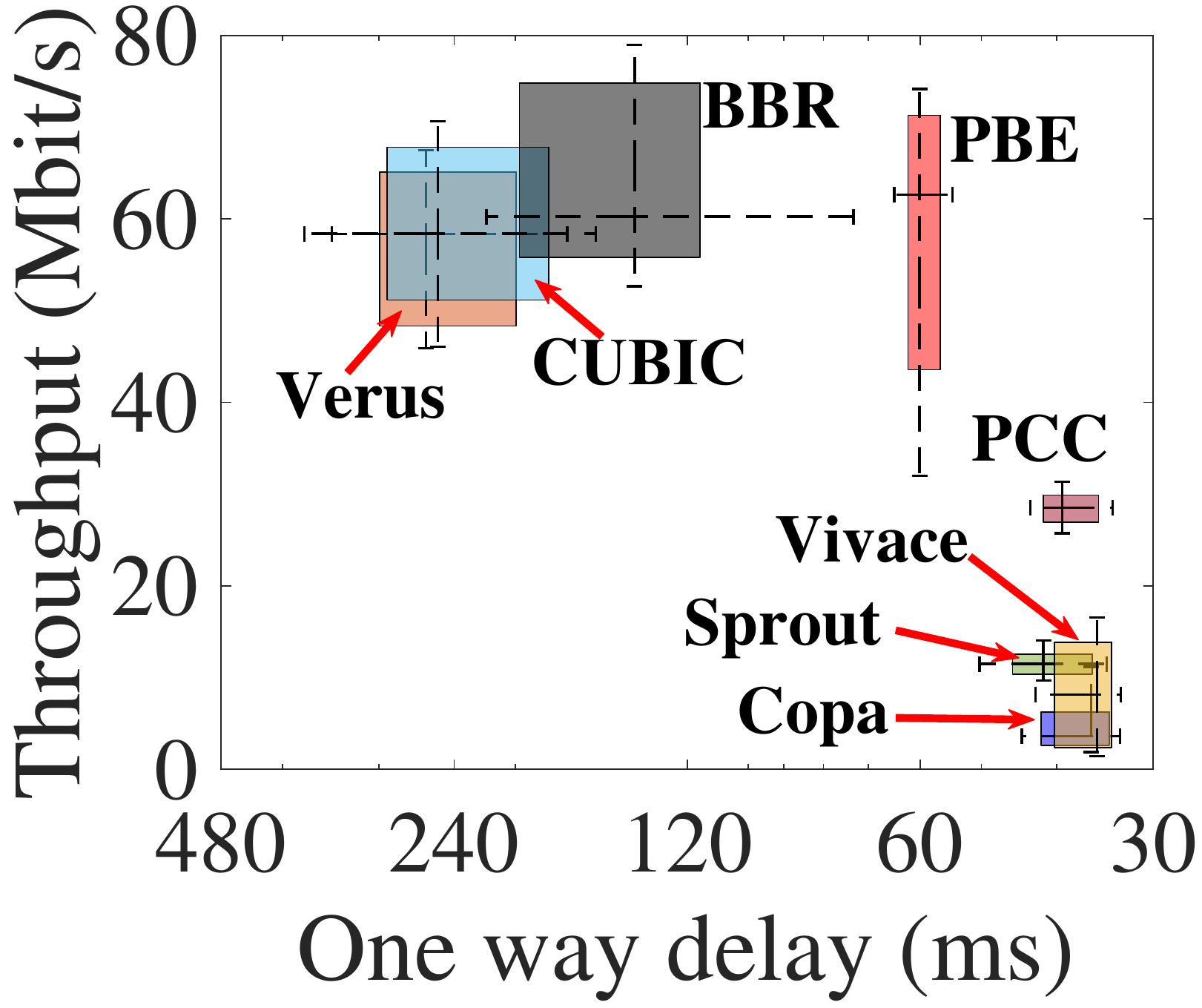}
        \caption{Achieved delay and throughput with controlled competing traffic.}
        \label{fig:CA_cell_rate}
    \end{minipage}
    \hfill
    \begin{minipage}[htb]{0.74\linewidth}
        \includegraphics[width=0.99\textwidth]{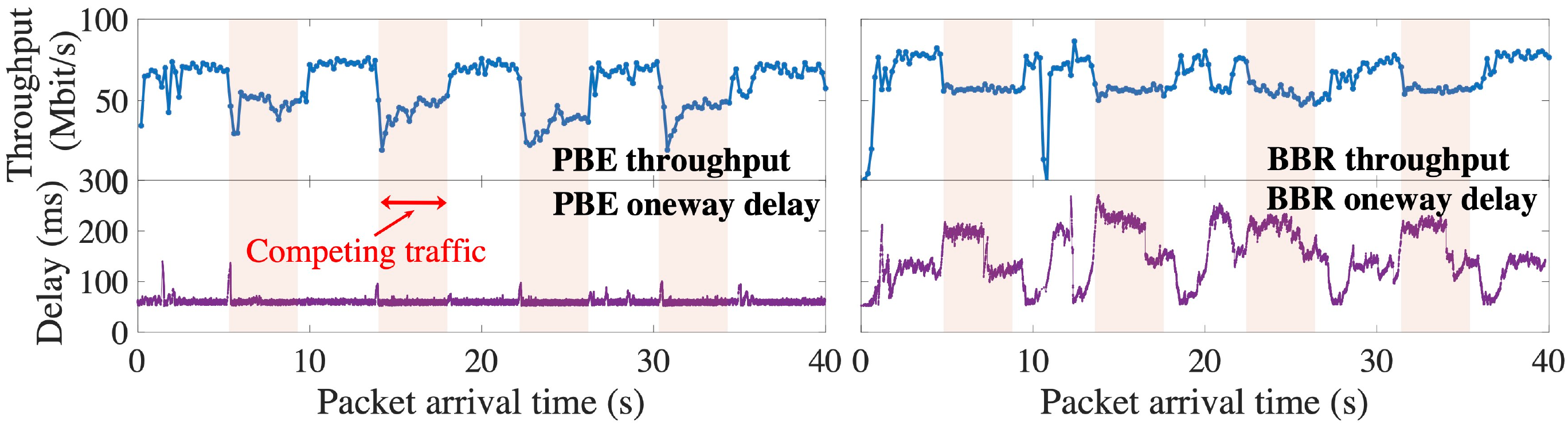}
      \caption{Average throughput and delay of every received packet in a flow.
      \systemnames{} rate increase and decrease is more responsive, thus grabbing capacity faster and keeping delay constant, respectively. In contrast, BBR suffers delay fluctuations.
      }
      \label{fig:channel_usage}
    \end{minipage}
    \vspace{-0.2cm}
\end{figure*}

\parahead{Alternation between states}
On average, \systemname{} spends 18\% and 4\% of its time working in Internet-bottleneck state, 
for 25 busy links and 15 idle links, respectively, 
which validates our assumption that a connection traversing a cellular network is bottlenecked at the cellular wireless link for most of the time.

\subsubsection{Performance under Mobility}
A major source of cellular wireless capacity variations arise from 
wireless channel quality variations, caused by client mobility. 
In this section, we investigate \systemnames{} performance under mobility. 
We conduct this experiment at night when the cell is approximately idle 
to reduce the capacity variations introduced by other random competing users. 
In each test, we put the phone at a location with RSSI of 
$-85$~dBm for the first 13 seconds,
and then move it along a predefined trajectory to another location with 
RSSI of $-105$~dBm in the next 13 seconds.
We move the phone back to the starting location ($-85$~dBm) 
with a faster speed, taking about four seconds and put it there for 10 seconds.
In total, each test takes 40 seconds.
We repeat the same process for each congestion control algorithm.

We present each algorithm's achieved throughput and delay in 
Figure~\ref{fig:mobility_delay_thput}, 
from which, we see that \systemname{} consistently achieves low delay 
($95^{\mathrm{th}}$ percentile of 64~ms) and high average 
throughput (55~Mbit/s).
BBR achieves comparable throughput (55~Mbit/s) with \systemname{} but 
suffers much higher delay (156~ms).
CUBIC and Verus achieve much lower throughput than 
\systemname{} (38~Mbit/s and 41~Mbit/s) and also introduces
high delay (296~ms and 467~ms). 
Other algorithms, \eg, PCC, PCC\hyp{}Vivace, Sprout, and Copa, have low 
throughput, resulting in under\hyp{}utilization of wireless capacity,
so mobility has a trivial effect on their packet delay. 

To further demonstrate \systemnames{} ability to track mobility, 
we divide the 40\hyp{}second experimentation period into 20 two\hyp{}second 
intervals and plot median throughput and delay of each interval 
for \systemname{} and BBR, in Figure~\ref{fig:mobility_BBR_PBE}.
We see that \systemname{} lowers and increases its send rate accurately 
when the signal strength decreases from 13 to 26 seconds and 
then increases from 26 to 30 seconds because of mobility,
resulting in nearly zero buffering in the network.
On the other hand, BBR overreacts to the signal strength decrease, 
reducing its send rate more than needed, because of its 
inaccurate end\hyp{}to\hyp{}end capacity estimation.
BBR also overestimates capacity when the signal quality 
recovers at 30 seconds, causing packet queuing 
and introducing excessive packet delay.

\subsubsection{Performance under Controlled Competition}
Besides mobility, 
the competition between mobile clients for limited wireless capacity 
is another major source of variations in network capacity.
In this section, we use controllable, on\hyp{}off
competing traffic to demonstrate \systemnames{} capability to track 
the time\hyp{}varying wireless bandwidth allocation caused by competition. 
Specifically, we start a \systemname{} flow that runs for 40 seconds 
using a Redmi~8 phone.  Every eight seconds, we also start a 
four\hyp{}second concurrent flow with a fixed offered load of 
60~Mbit/s from an AWS server, using a Xiaomi MIX3.
We conduct the experiments at night to make the possibility 
of uncontrolled competition from other users remote. 
We repeat the experiment using different congestion control algorithms. 

We plot each algorithm's throughput and delay in Figure \ref{fig:CA_cell_rate}, 
from which, we see that only \systemname{} can simultaneously 
achieve high throughput and low latency.
The average throughput of \systemname{} is 57~Mbit/s, comparable with 
CUBIC at 58~Mbit/s, and Verus at 56~Mbit/s, but slightly smaller than 
BBR at 62~Mbit/s. But the average and $95^{\mathrm{th}}$ 
percentile delay of \systemname{} is 61~ms and 71~ms, much smaller than 
BBR at 147~ms and 227~ms, CUBIC at 252~ms and 416~ms, and 
Verus at 263~ms and 403~ms. 
To further demonstrate \systemnames{} and BBR's reactions to competing traffic, 
we also plot the throughput (averaged over every 200~millisecond interval) 
and the delay of all received packets, in Figure~\ref{fig:channel_usage}, 
where the shaded areas represent the time periods
when the concurrent competing traffic generated by the MIX3 is present.
We see that \systemname{} accurately tracks the entrance of the competitor 
and lowers its sending rate promptly, resulting in nearly no packet queuing. 
\systemname{} immediately grabs the idle bandwidth when the competing traffic finishes its flow, maximizing the achieved throughput. 
In contrast, BBR cannot timely detects the decreasing capacity caused by competing traffic, resulting significantly enlarged delay.
\begin{figure}
    \centering
        \begin{subfigure}[b]{0.49\linewidth}
            \centering
            \includegraphics[width=0.99\textwidth]{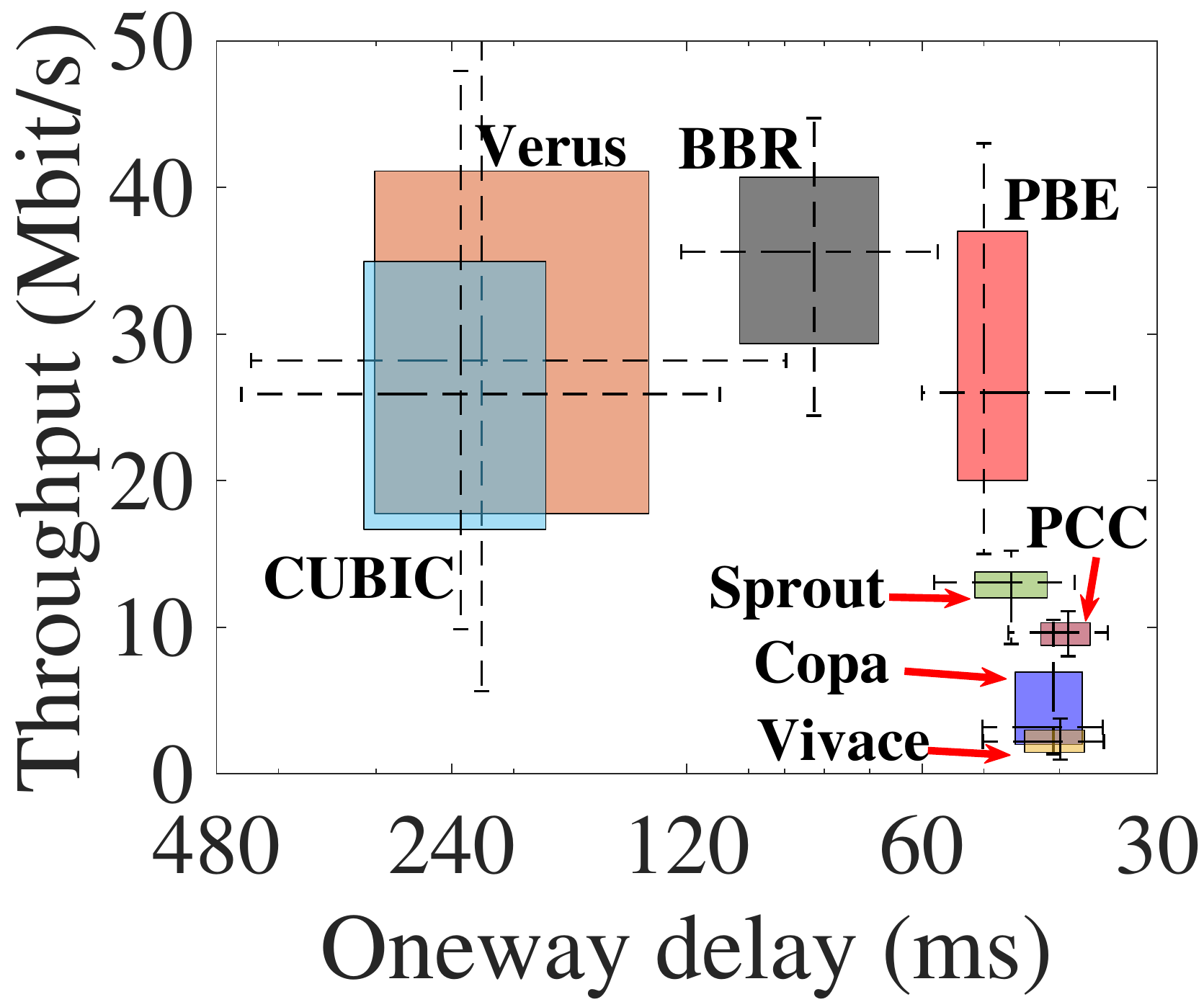}
            \caption{The first connection.}
            \label{fig:2flow_flow1}
        \end{subfigure}
        \hfill
        \begin{subfigure}[b]{0.49\linewidth}
            \centering
            \includegraphics[width=0.99\textwidth]{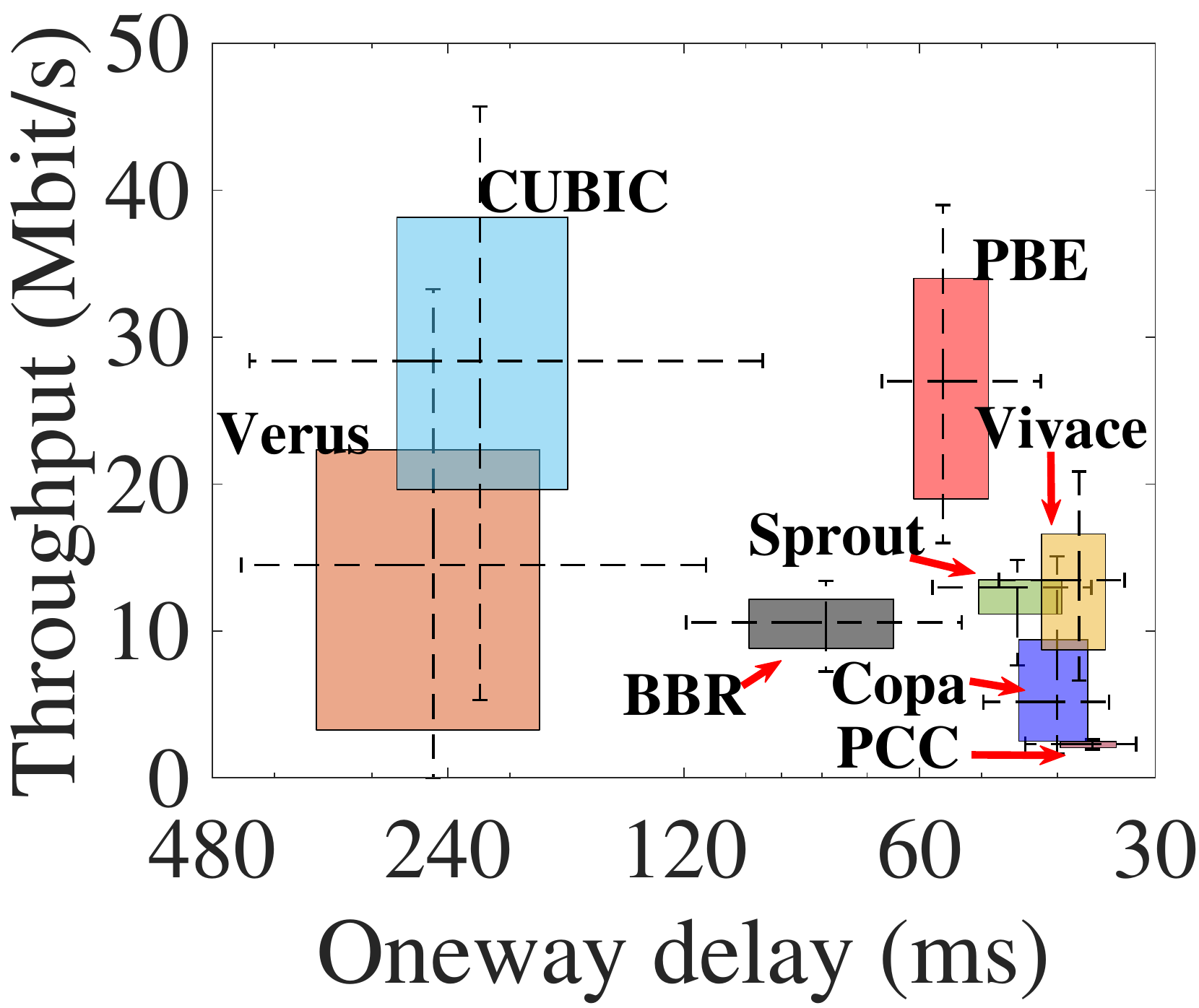}
            \caption{The second connection.}
            \label{fig:2flow_flow2}
        \end{subfigure}
        \caption{The oneway delay and throughput achieved by eight congestion control algorithms for two concurrent connections between one device and two remote servers.}
        \label{fig:delay_thput_2flow}
        \vspace{-0.4cm}
\end{figure}

\begin{figure*}[tbh]
    \centering
    \begin{subfigure}[b]{0.24\linewidth}
        \centering
        \includegraphics[width=0.99\textwidth]{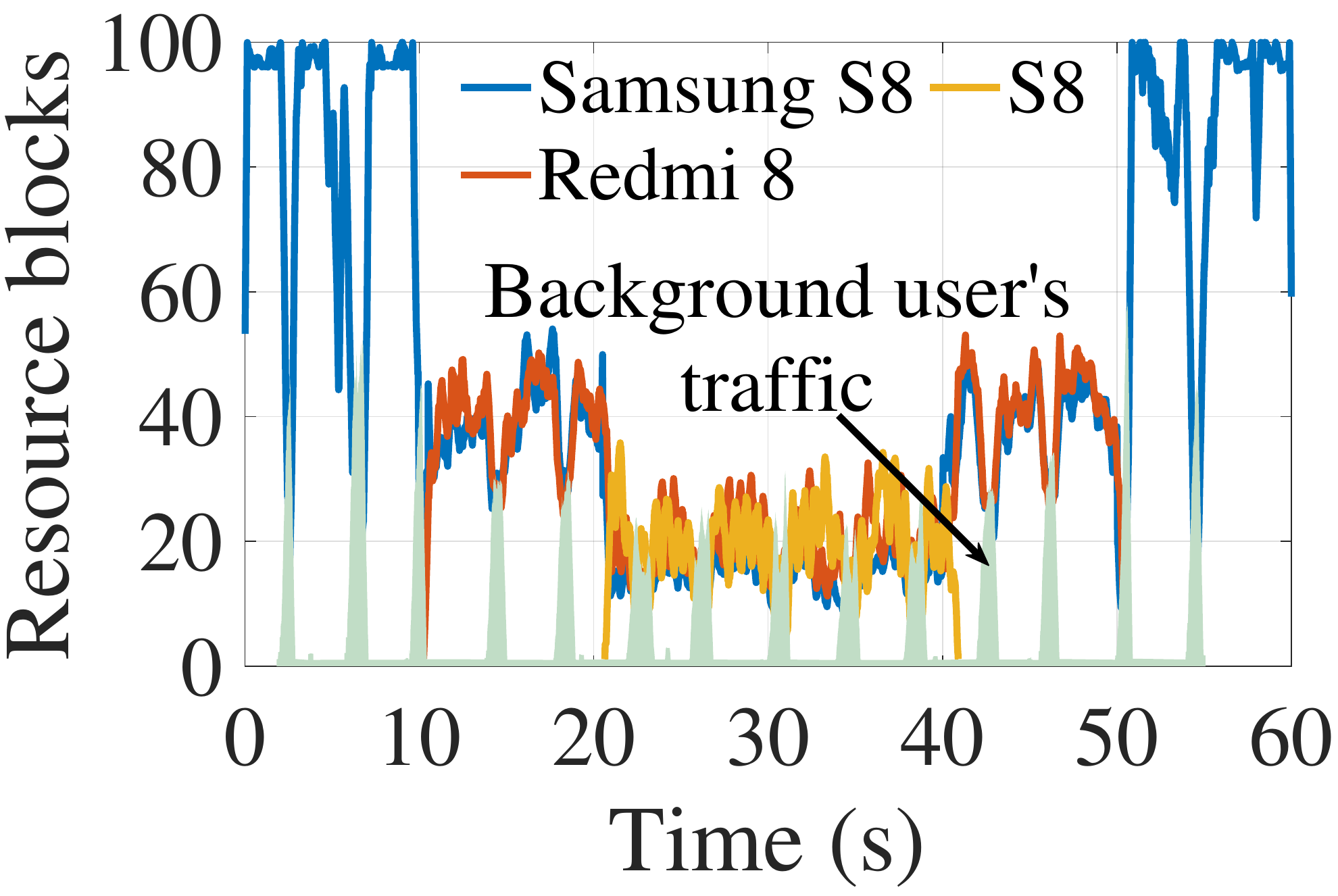}
        \caption{Three \systemname{} flows with similar RTTs.}
        \label{fig:fair_multi}
    \end{subfigure}
    \hfill
    \begin{subfigure}[b]{0.24\linewidth}
        \centering
        \includegraphics[width=0.99\textwidth]{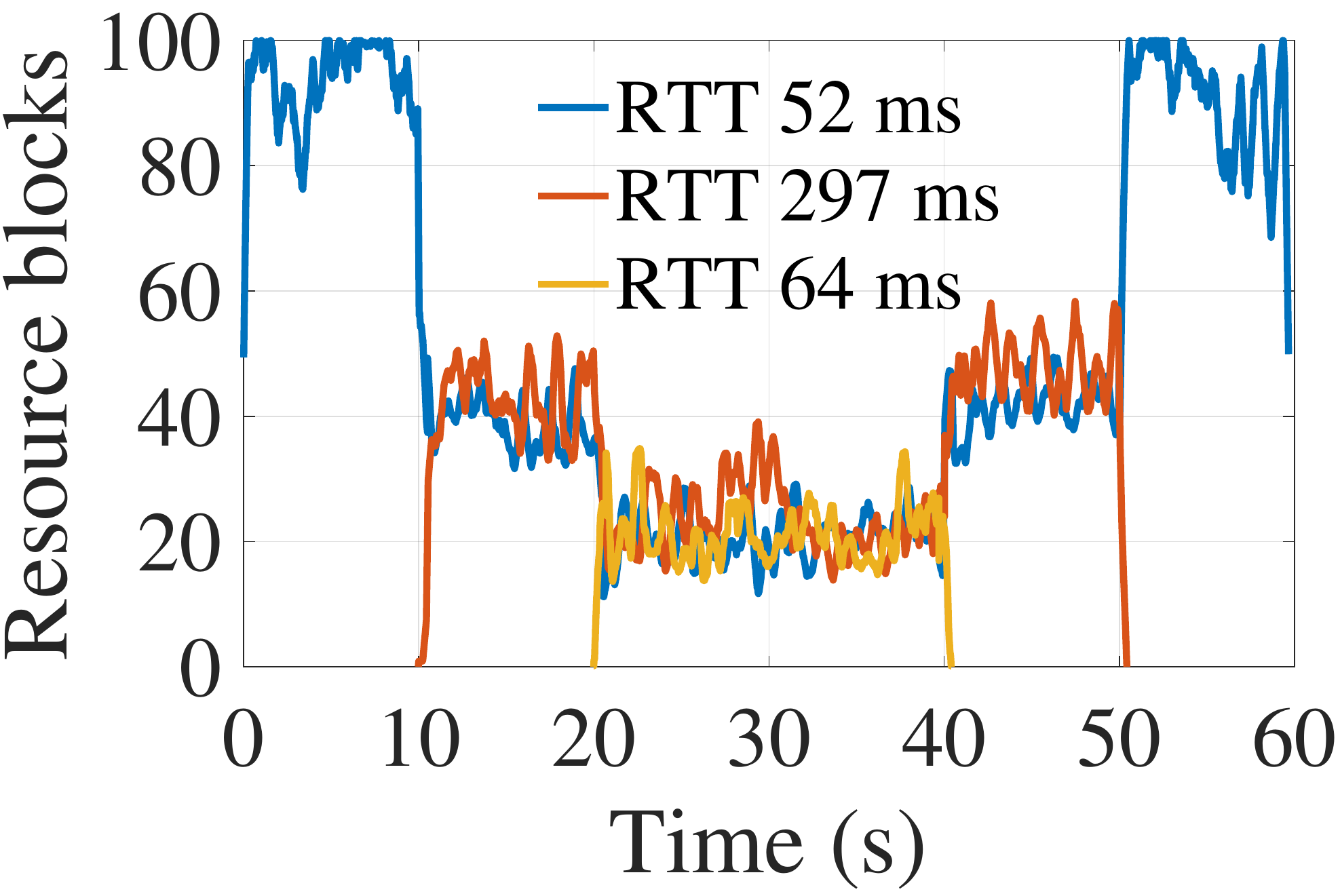}
        \caption{Three \systemname{} flows with significant RTT differences.}
        \label{fig:fair_rtt}
    \end{subfigure}
    \hfill
    \begin{subfigure}[b]{0.24\linewidth}
        \centering
        \includegraphics[width=0.99\textwidth]{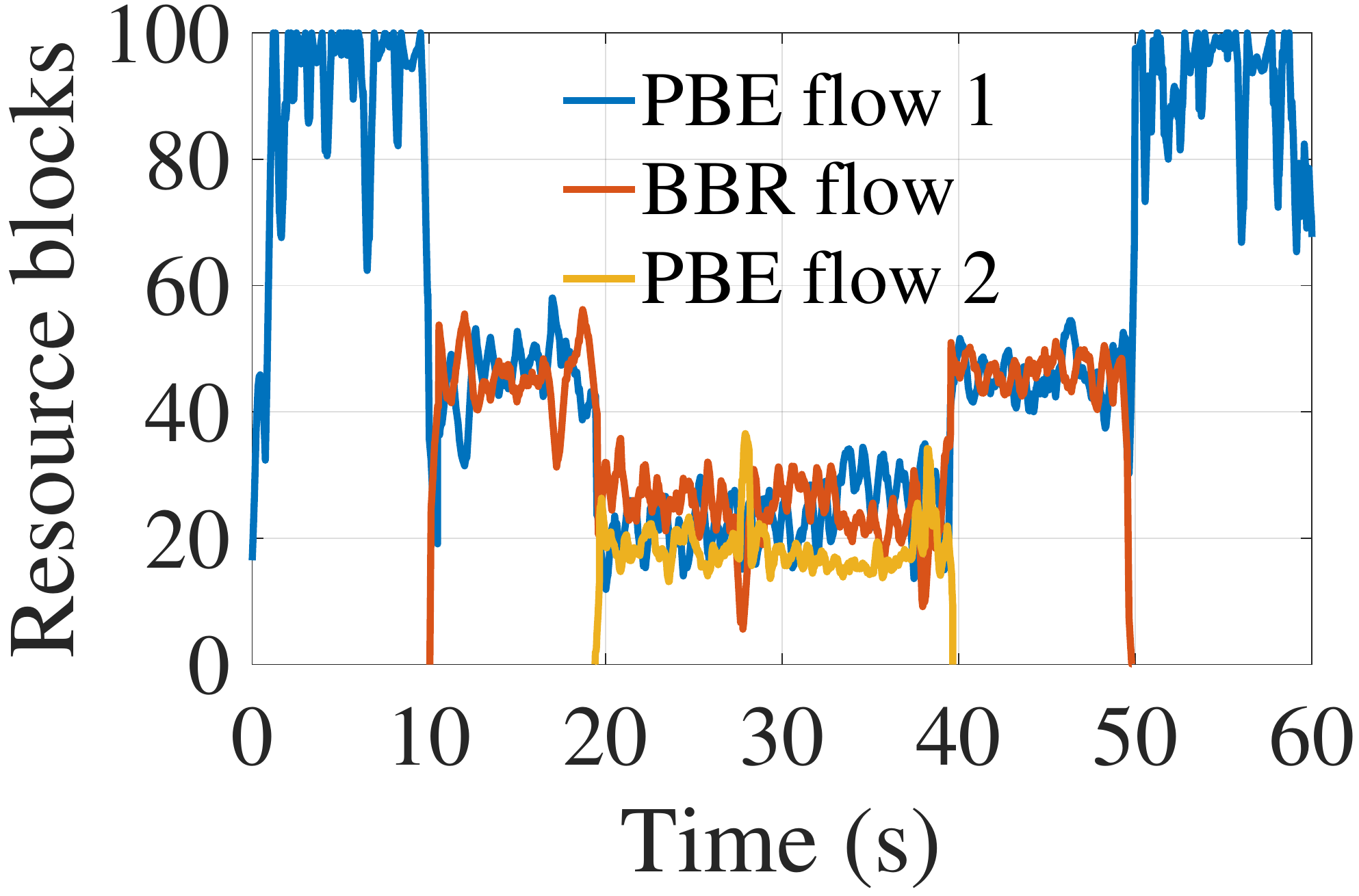}
        \caption{Two \systemname{} flows coexist with one BBR flow.}
        \label{fig:fair_bbr}
    \end{subfigure}
    \hfill
    \begin{subfigure}[b]{0.24\linewidth}
        \centering
        \includegraphics[width=0.99\textwidth]{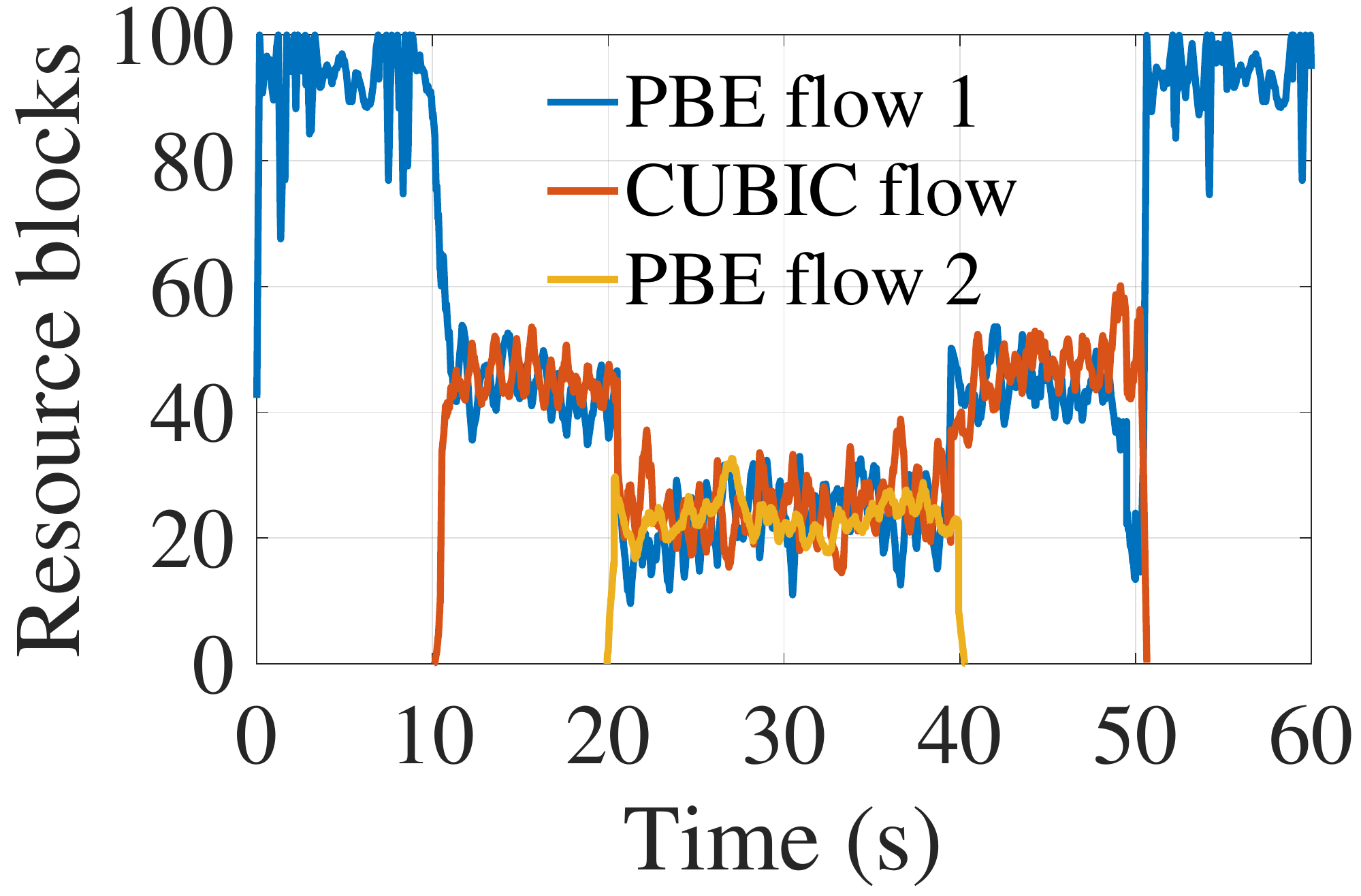}
        \caption{Two \systemname{} flows coexist with one CUBIC flow.}
        \label{fig:fair_cubic}
    \end{subfigure}
    \caption{The allocated PRBs (averaged over 50 subframes) by the primary cell to three mobile phones, when these three mobile phones starts three \systemname{} flows with three AWS servers in US \textbf{(a)};  three \systemname{} flows with two AWS servers in US and one AWS server in Singapore \textbf{(b)}; two \systemname{} flows with one BBR flow \textbf{(c)}; two \systemname{} flows with one CUBIC flow \textbf{(d)}.
    }
    \label{fig:fairness}
    \vspace{-0.2cm}
\end{figure*}

\subsubsection{Single device multiple connections}
In this section, we evaluate how \systemname{} performs in the scenario 
where one device simultaneously starts multiple connections with different remote servers. 
Specifically, we let the MIX3 start two concurrent flows with two AWS servers, each running for 40 seconds. 
We repeat the experiments using different congestion control algorithms, 
and plot each algorithm's throughput and delay in Figure~\ref{fig:delay_thput_2flow}.
We see that \systemname{} achieves high throughput and low delay for both flows.
The average throughput is 26~Mbit/s and 28~Mbit/s, and the median delay is 48~ms and 56~ms, for the first and second flow, respectively.
Furthermore, \systemname{} fairly allocates the estimated capacity for two flows 
so these two flows have similar throughput, 
while other algorithms may result in unbalanced throughput for multiple flows,
\eg, BBR achieves 10~Mbit/s and 35~Mbit/s for the first and second connection, respectively.
We note that even though \systemname{} may achieve a smaller throughput for a single connection compared to other algorithms, \eg, the first connection comparing with BBR,
\systemname{} provides better fairness across connections. 

\subsection{Fairness}\label{s:eval:fair}
In this section, we evaluate the fairness of \systemname{}, 
focusing on the case where the bottleneck is the cellular wireless link.

\parahead{Methodology}
Without knowing the base station's 
resource allocation algorithm and fairness policy, 
simulation\hyp{}based experiments cannot predict real\hyp{}world cellular network behavior.
We therefore, evaluate \systemnames{} fairness directly 
in a cellular deployment.
To eliminate the impact of background traffic, 
we conduct our experiment at night when the cell is idle.
We use the three phones as three competing users, 
each setting up a connection with a AWS server.  
The S8, Redmi~8 and MIX3 starts its flow at zero, 10, and 20 seconds, 
and ends at 60, 50, and 40 seconds, respectively.
These three phones share the same primary cell but have 
different secondary and tertiary cells (if configured), 
so the primary cell at 1.94~GHz is the shared bottleneck of three connections. 
We record the allocated PRBs to each user by the primary cell, when three connections are running concurrently.
Three connections get identical allocated primary cell PRBs 
if they achieve a fair\hyp{}share. 

\subsubsection{Multi-user fairness}\label{s:fair_multi}
We investigate the fairness between multiple \systemname{} 
flows with similar propagation delays.
We setup three AWS servers in the US and start three current connections 
via three mobile phones, plotting the allocated bandwidth by 
the primary cell to the three phones in Figure~\ref{fig:fair_multi}. 
We see that the \systemname{} flows quickly converge to the fair\hyp{}share
of the bottleneck bandwidth. 
Jain's fairness index \cite{JainIndex} is 99.97 and 98.73\% with two 
and three concurrent flows (100\% is ideal), respectively.
Since we cannot prevent all associated users from using the 
cellular network, we observe light background traffic generated by 
a unknown user, in this experiment.
The \systemname{} flow also reacts quickly, fairly sharing the bandwidth 
with background users.

\subsubsection{RTT fairness}\label{s:fair_RTT}
We investigate whether \systemname{} can guarantee a fair\hyp{}share 
of wireless link capacity between multiple flows with 
significant differences in propagation delay.
We use three mobile phones to build concurrent connections 
with three AWS servers: one in Singapore (average RTT of 297~ms) and two in the 
US (average RTTs of 52~ms and 64~ms).
We plot the the primary cell allocated PRBs for these
connections in Figure~\ref{fig:fair_rtt}. 
We see that the all three \systemname{} flows with significant 
propagation delay differences obtain similar allocated bandwidths.
Jain's fairness indices are 99.74\% and 99.45\% with two 
and three concurrent flows, respectively.

\subsubsection{TCP friendliness}\label{s:fair_TCP}
A common requirement from new congestion control schemes is
the capability of fairly sharing the available bandwidth with existing congestion control algorithms like BBR and CUBIC.
We investigate the performance of \systemname{} 
in two cases: two \systemname{} flows coexisting
with one BBR flow, and two \systemname{} flows coexisting with
one CUBIC flow.
Figures~\ref{fig:fair_bbr} and~\ref{fig:fair_cubic} depict 
allocated PRBs for three connections in these cases, 
showing that \systemname{} shares bottleneck bandwidth equally 
with both CUBIC and BBR flows. 
Jain's fairness index is 99.96\% and 98.52\% with two and three concurrent 
flows in Figure~\ref{fig:fair_bbr}, 
and 99.95\% and 98.34\% with two and three flows in Figure~\ref{fig:fair_cubic}.
The base station fairness policy prevents one user from grabbing all the bandwidth. 
Though CUBIC and BBR may aggressively increase their sending rate, 
the base station limits the total bandwidth they 
can obtain and forces them to share with other concurrent flows. 
\section{Discussion}
\parahead{Power consumption}
In the connected state, 
a mobile device must keep its radio on and decodes the control channel to check whether the base station has data for it or not in each subframe. 
Therefore, \systemname{} does not turn the radio of mobile device on for any extra time than necessary currently and thus introduces no additional power costs. 
The small computational overhead PBE-CC introduces is that the mobile device may need to decode control messages that are not transmitted to it. 
But, the number of extra control messages inside each subframe the device needs to decode is very small, since 
our experimental results shows that there are less than 4 control messages inside more than 95\% subframes. 
Furthermore, the control messages are very short (less than 70 bits), 
so that decoding one message only involves small extra computational overhead. 

\parahead{Packet buffering}
\systemname{} works at or very close to the Kleinrock TCP operating point~\cite{kpoint1,kpoint2}, 
which minimizes buffering, minimizing the delay.
In practice, it could be beneficial to buffer some bytes in the base station, 
which slightly increases delay but helps to immediately utilize
increases in connection throughput, 
before the sender modulates its sending rate (congestion control has at least a 
round trip time delay).
In the future, we plan to extend \systemname{} to enable the sender\fshyp{}app 
to adaptively adjust the buffering inside the network, 
trading off increased delay for increased throughput.

\parahead{Fairness policy}
Currently, \systemname{} fairly shares 
idle bandwidth among all active users in the connection start state. 
In the future, \systemname{} can be modified to incorporate other fairness policies, 
\eg, active users with lower physical data rate grab larger bandwidth. 
\systemnames{} control algorithm adapts to an arbitrary fairness policy,
achieving equilibrium in the steady state.

\parahead{Misreported congestion feedback}
\systemname{} relies on the mobile user to report the estimated capacity back to the server 
so it is possible that a malicious user may report a data rate higher than the network can support, 
triggering overwhelming number of data being injected into the network, causing catastrophic impact. 
In future work, \systemname{} can be extended to 
detect such malicious users via implementing a server side BBR-like throughput estimator, 
which estimates the currently achieved throughput purely with timestamps of packets being sent and acknowledged, 
without any involvement of the mobile user. 
By comparing the achieved throughput and capacity reported by the user, 
\systemname{} identifies any user who consistently 
reports a rate higher than the achievable throughput as a malicious user.
\section{Conclusion}
\label{s:concl}

\systemname{} is the first end\hyp{}to\hyp{}end congestion control 
algorithm to seamlessly integrate mobile client\hyp{}side 
wireless physical layer capacity measurement into its design,
which is crucial for the multi\hyp{}cell design
of 4G and 5G wireless networks.  Our rigorous performance evaluation
featuring multi\hyp{}locations, mobility, varying background traffic
levels, and varying RTTs shows that \systemname{} outperforms 
many leading congestion control 
algorithms in both latency and throughput.  \systemname{} is
also immediately deployable, with modifications required solely to
content servers and  mobile clients.
This work does not raise any ethical issues.
\section*{Acknowledgements}
We thank the anonymous SIGCOMM reviewers and our shepherd for their valuable 
feedback that has improved the quality of this paper.
This work was supported by NSF grant CNS-1617161.
\clearpage
\pagenumbering{roman}
\bibliographystyle{concise2}
\begin{raggedright}
\renewcommand{\bibfont}{\normalsize}
\bibliography{references}
\end{raggedright}
\end{document}